# *GeT_Move*: An Efficient and Unifying Spatio-Temporal Pattern Mining Algorithm for Moving Objects

Phan Nhat Hai
LIRMM - Univ. Montpellier 2
nhat-hai.phan@teledetection.fr

Pascal Poncelet
LIRMM - Univ. Montpellier 2
pascal.poncelet@lirmm.fr

Maguelonne Teisseire
TETIS - Montpellier, France
teisseire@teledetection.fr

## ABSTRACT

Recent improvements in positioning technology has led to a much wider availability of massive moving object data. A crucial task is to find the moving objects that travel together. Usually, these object sets are called spatio-temporal patterns. Due to the emergence of many different kinds of spatio-temporal patterns in recent years, different approaches have been proposed to extract them. However, each approach only focuses on mining a specific kind of pattern. In addition to being a painstaking task due to the large number of algorithms used to mine and manage patterns, it is also time consuming. Moreover, we have to execute these algorithms again whenever new data are added to the existing database. To address these issues, we first redefine spatio-temporal patterns in the itemset context. Secondly, we propose a unifying approach, named *GeT_Move*, which uses a frequent closed itemset-based spatio-temporal pattern-mining algorithm to mine and manage different spatio-temporal patterns. GeT_Move is implemented in two versions which are GeT_Move and Incremental GeT_Move. To optimize the efficiency and to free the parameters setting, we also propose a Parameter Free Incremental GeT_Move algorithm. Comprehensive experiments are performed on real datasets as well as large synthetic datasets to demonstrate the effectiveness and efficiency of our approaches.

## 1. INTRODUCTION

Nowadays, many electronic devices are used for real world applications. Telemetry attached on wildlife, GPS installed in cars, sensor networks, and mobile phones have enabled the tracking of almost any kind of data and has led to an increasingly large amount of data that contain moving objects and numerical data. Therefore, analysis on such data to find interesting patterns is attracting increasing attention for applications such as movement pattern analysis, animal behavior study, route planning and vehicle control.

Early approaches designed to recover information from spatio-temporal datasets included ad-hoc queries aimed as answering queries concerning a single predicate range or nearest neighbour. For instance, *"finding all the moving objects inside area A between 10:00 am and 2:00 pm"* or *"how many cars were driven between Main Square and the Airport on Friday"* [29]. Spatial query extensions in GIS applications are able to run this type of query. However, these techniques are used to find the best solution by exploring each spatial object at a specific time according to some metric distance measurement (usually Euclidean). As results, it is difficult to capture collective behaviour and correlations among the involved entities using this type of queries.

Recently, there has been growing interest in the querying of patterns which capture *'group'* or *'common'* behaviour among moving entities. This is particularly true for the development of approaches to identify groups of moving objects for which a strong relationship and interaction exist within a defined spatial region during a given time duration. Some examples of these patterns are flocks [1, 2, 14, 15], moving clusters [4, 18, 7], convoy queries [3, 16], stars and k-stars [17], closed swarms [6, 13], group patterns [21], periodic patterns [25], co-location patterns [22], TraLus [23], etc...

To extract these kinds of patterns, different algorithms have been proposed. Naturally, the computation is costly and time consuming because we need to execute different algorithms consecutively. However, if we had an algorithm which could extract different kinds of patterns, the computation costs will be significantly decreased and the process would be much less time consuming. Therefore, we need to develop an efficient unifying algorithm.

In some applications (e.g. cars), object locations are continuously reported by using Global Positioning System (GPS). Therefore, new data is always available. If we do not have an incremental algorithm, we need to execute again and again algorithms on the whole database including existing data and new data to extract patterns. This is of course, cost-prohibitive and time consuming. An incremental algorithm can indeed improve the process by combining the results extracted from the existing data and the new data to obtain the final results.

With the above issues in mind, we propose *GeT_Move*: a unifying incremental spatio-temporal pattern-mining approach. Part of this approach is based on an existing state-of-the-art algorithm which is extended to take advantage of well-known frequent closed itemset mining algorithms. In order to use it, we first redefine spatio-temporal patterns in an itemset context. Secondly, we propose a spatio-temporal matrix to describe original data and then an incremental fre-quent closed itemset-based spatio-temporal pattern-mining algorithm to extract patterns.

Naturally, obtaining the optimal parameters is a difficult task for most of algorithms which require parameters setting. Even if we are able to obtain the optimal parameters after doing many executions and evaluate the results on a dataset, the optimal values of parameters will be different on the other datasets. To tackle this issue and to optimize the efficiency as well as to free the parameters setting, we propose a parameter free incremental GeT_Move. The main idea is to re-arrange the input data based on nested concept [31] so that incremental GeT_Move can automatically extract patterns without parameters setting efficiently.

The main contributions of this paper are summarized below.

- We re-define the spatio-temporal patterns mining in the itemset context which enable us to effectively extract different kinds of spatio-temporal patterns.
- We present incremental approaches, named *GeT_Move* and *Incremental GeT_Move*, which efficiently extract frequent closed itemsets from which spatio-temporal patterns are retrieved.
- We design and propose a parameter free incremental GeT_Move. The advantages of this approach is that it does not require the use to set parameters and automatically extract patterns efficiently.
- We propose to deal with new data arriving and propose an explicit combination of pairs of frequent closed itemsets-based pattern mining algorithm which efficiently combines the results in the existing database with the arriving data to obtain the final results.
- We present comprehensive experimental results over both real and synthetic databases. The results demonstrate that our techniques enable us to effectively extract different kinds of patterns. Furthermore, our approaches are more efficient compared to other algorithms in most of cases.

The remaining sections of the paper are organized as follows. Section 2 discusses preliminary definitions of the spatio-temporal patterns as well as the related work. The patterns such as swarms, closed swarms, convoys and group patterns are redefined in an itemset context in Section 3. We present our approaches in Section 4. Experiments testing effectiveness and efficiency are shown in Section 5. Finally, we draw our conclusions in Section 6.

# 2. SPATIO-TEMPORAL PATTERNS

In this section we briefly propose an overview of the main spatio-temporal patterns. We thus define the different kinds of patterns and then we discuss the related work.

## 2.1 Preliminary Definitions

The problem of spatio-temporal patterns has been extensively addressed over the last years. Basically, a spatio-temporal patterns are designed to group similar trajectories or objects which tend to move together during a time interval. So many different definitions can be proposed and today lots of patterns have been defined such as flocks [1, 2, 14, 15], convoys [3, 16], swarms, closed swarms [6, 13], moving clusters [4, 18, 7] and even periodic patterns [25].

In this paper, we focus on proposing a unifying approach to effectively and efficiently extract all these different kinds of patterns. First of all, we assume that we have a group of moving object $O_{DB} = \{o_1, o_2, \ldots, o_z\}$, a set of timestamps $T_{DB} = \{t_1, t_2, \ldots, t_n\}$ and at each timestamp $t_i \in T_{DB}$, spatial information[1] $x, y$ for each object. For example, Table 1 illustrates an example of a spatio-temporal database. Usually, in spatio-temporal mining, we are interested in extracting a group of objects staying together during a period. Therefore, from now, $O = \{o_{i_1}, o_{i_2}, \ldots, o_{i_p}\} (O \subseteq O_{DB})$ stands for a group of objects, $T = \{t_{a_1}, t_{a_2}, \ldots, t_{a_m}\} (T \subseteq T_{DB})$ is the set of timestamps within which objects stay together. Let $\varepsilon$ be the user-defined threshold standing for minimum number of objects and $min_t$ the minimum number of timestamps. Thus $|O|$ (resp. $|T|$) must be greater than or equal to $\varepsilon$ (resp. $min_t$).

**Table 1: An example of a Spatio-Temporal Database**

| Objects $O_{DB}$ | Timesets $T_{DB}$ | $x$ | $y$ |
|---|---|---|---|
| $o_1$ | $t_1$ | 2.3 | 1.2 |
| $o_2$ | $t_1$ | 2.1 | 1 |
| $o_1$ | $t_2$ | 10.3 | 28.1 |
| $o_2$ | $t_2$ | 0.3 | 1.2 |

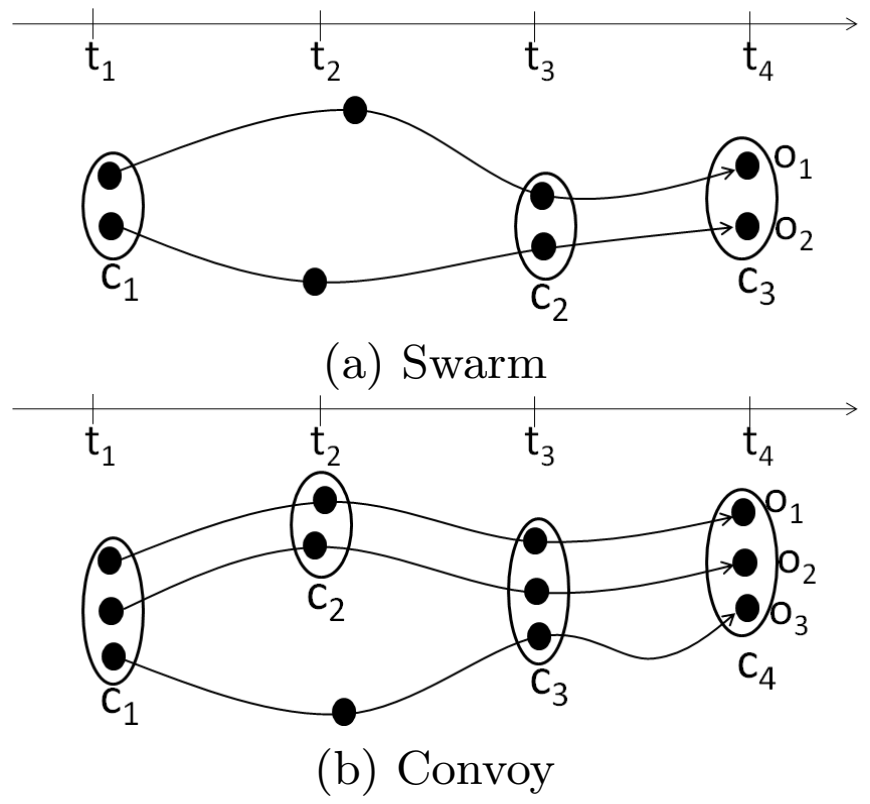

**Figure 1: An example of swarm and convoy where $c_1, c_2, c_3, c_4$ are clusters gathering closed objects together at specific timestamps.**

In the following, we formally define all the different kinds of patterns.

Informally, a *swarm* is a group of moving objects $O$ containing at least $\varepsilon$ individuals which are closed each other for at least $min_t$ timestamps. Then a swarm can be formally defined as follows:

Definition 1. *Swarm [6]. A pair $(O, T)$ is a swarm if:*

$$\begin{cases} (1): \forall t_{a_i} \in T, \exists c \; s.t. \; O \subseteq c, \; c \text{ is a cluster.} \\ \text{There is at least one cluster containing} \\ \text{all the objects in } O \text{ at each timestamp in } T. \\ (2): |O| \geq \varepsilon. \\ \text{There must be at least } \varepsilon \text{ objects.} \\ (3): |T| \geq min_t. \\ \text{There must be at least } min_t \text{ timestamps.} \end{cases} \quad (1)$$

For example, as shown in Figure 1a, if we set $\varepsilon = 2$ and $min_t = 2$, we can find the following swarms

[1]Spatial information can be for instance GPS location.

$(\{o_1, o_2\}, \{t_1, t_3\})$, $(\{o_1, o_2\}, \{t_1, t_4\})$, $(\{o_1, o_2\}, \{t_3, t_4\})$, $(\{o_1, o_2\}, \{t_1, t_3, t_4\})$. We can note that these swarms are in fact redundant since they can be grouped together in the following swarm $(\{o_1, o_2\}, \{t_1, t_3, t_4\})$.

To avoid this redundancy, Zhenhui Li et al. [6] propose the notion of *closed swarm* for grouping together both objects and time. A swarm $(O, T)$ is *object-closed* if, when fixing $T$, $O$ cannot be enlarged. Similarly, a swarm $(O, T)$ is *time-closed* if, when fixing $O$, $T$ cannot be enlarged. Finally, a swarm $(O, T)$ is a closed swarm if it is both object-closed and time-closed and can be defined as follows:

DEFINITION 2. *Closed Swarm [6]. A pair $(O, T)$ is a closed swarm if:*

$$\begin{cases} (1) : (O, T) \text{ is a swarm.} \\ (2) : \nexists O' \text{ s.t. } (O', T) \text{ is a swarm and } O \subset O'. \\ (3) : \nexists T' \text{ s.t. } (O, T') \text{ is a swarm and } T \subset T'. \end{cases} \quad (2)$$

For instance, in the previous example, $(\{o_1, o_2\}, \{t_1, t_3, t_4\})$ is a closed swarm.

A *convoy* is also a group of objects such that these objects are closed each other during at least $min_t$ time points. The main difference between convoy and swarm (or closed swarm) is that convoy lifetimes must be consecutive. In essential, by adding the consecutiveness condition to swarms, we can define convoy as follows:

DEFINITION 3. *Convoy [3]. A pair $(O, T)$, is a convoy if:*

$$\begin{cases} (1) : (O, T) \text{ is a swarm.} \\ (2) : \forall i, 1 \leq i < |T|, \; t_{a_i}, t_{a_{i+1}} \text{ are consecutive.} \end{cases} \quad (3)$$

For instance, on Figure 1b, with $\varepsilon = 2, min_t = 2$ we have two convoys $(\{o_1, o_2\}, \{t_1, t_2, t_3, t_4\})$ and $(\{o_1, o_2, o_3\}, \{t_3, t_4\})$.

Until now, we have considered that we have a group of objects that move close to each other for a long time interval. For instance, as shown in [28], moving clusters and different kinds of flocks virtually share essentially the same definition. Basically, the main difference is based on the clustering techniques used. Flocks, for instance, usually consider a rigid definition of the radius while moving clusters and convoys apply a density-based clustering algorithm (e.g. DBScan [5]). Moving clusters can be seen as special cases of convoys with the additional condition that they need to share some objects between two consecutive timestamps [28]. Therefore, in the following, for brevity and clarity sake we will mainly focus on convoy and density-based clustering algorithm.

According to the previous definitions, the main difference between convoys and swarms is about the consecutiveness and non-consecutiveness of clusters during a time interval. In [21], Hwang et al. propose a general pattern, called a *group pattern*, which essentially is a combination of both convoys and closed swarms. Basically, group pattern is a set of disjointed convoys which are generated by the same group of objects in different time intervals. By considering a convoy as a timepoint, a group pattern can be seen as a swarm of disjointed convoys. Additionally, group pattern cannot be enlarged in terms of objects and number of convoys. Therefore, group pattern is essentially a closed swarm of disjointed convoys. Formally, group pattern can be defined as follows:

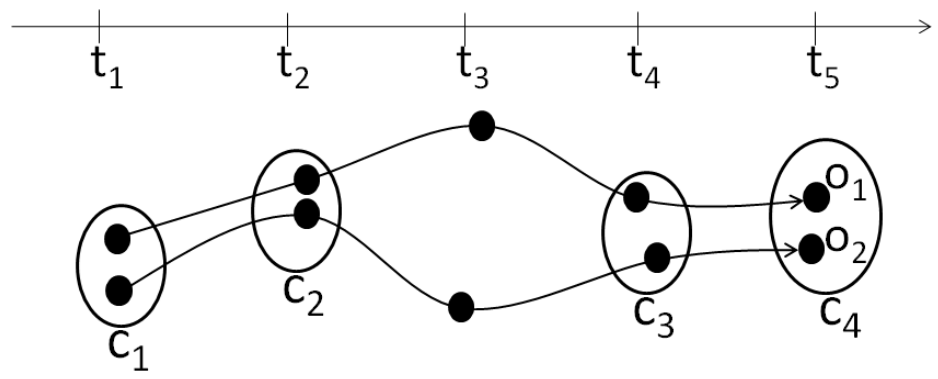

**Figure 2: A group pattern example.**

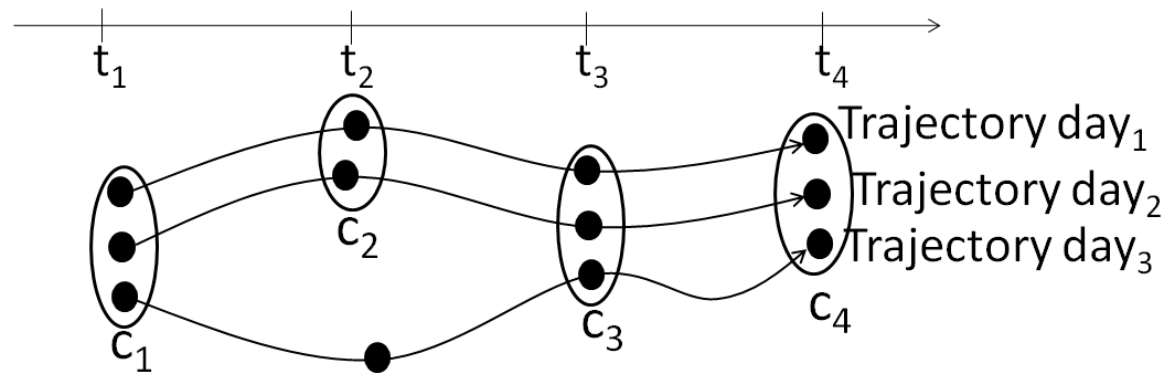

**Figure 3: A periodic pattern example.**

DEFINITION 4. *Group Pattern [21]. Given a set of objects $O$, a minimum weight threshold $min_{wei}$, a set of disjointed convoys $T_{\mathcal{S}} = \{s_1, s_2, \ldots, s_n\}$, a minimum number of convoys $min_c$. $(O, T_{\mathcal{S}})$ is a group pattern if:*

$$\begin{cases} (1) : (O, T_{\mathcal{S}}) \text{ is a closed swarm with } \varepsilon, min_c. \\ (2) : \frac{\sum_{i=1}^{|T_{\mathcal{S}}|} |s_i|}{|T_{DB}|} \geq min_{wei}. \end{cases} \quad (4)$$

*Note that $min_c$ is only applied for $T_{\mathcal{S}}$ (e.g. $|T_{\mathcal{S}}| \geq min_c$).*

For instance, see Figure 2, with $min_t = 2$ and $\varepsilon = 2$ we have a set of convoys $T_{\mathcal{S}} = \{(\{o_1, o_2\}, \{t_1, t_2\}), (\{o_1, o_2\}, \{t_4, t_5\})\}$. Additionally, with $min_c = 1$ we have $(\{o_1, o_2\}, T_{\mathcal{S}})$ is a closed swarm of convoys because $|T_{\mathcal{S}}| = 2 \geq min_c$, $|O| \geq \varepsilon$ and $(O, T_{\mathcal{S}})$ cannot be enlarged. Furthermore, with $min_{wei} = 0.5$, $(O, T_{\mathcal{S}})$ is a group pattern since $\frac{|[t_1, t_2]| + |[t_4, t_5]|}{|T_{DB}|} = \frac{4}{5} \geq min_{wei}$.

Previously, we overviewed patterns in which group objects move together during some time intervals. However, mining patterns from individual object movement is also interesting. In [25], N. Mamoulis et al. propose the notion of *periodic patterns* in which an object follows the same routes (approximately) over regular time intervals. For example, people wake up at the same time and generally follow the same route to their work everyday. Informally, given an object's trajectory including $\mathcal{N}$ timepoints, $\mathcal{T}_{\mathcal{P}}$ which is the number of timestamps that a pattern may re-appear. An object's trajectory is decomposed into $\lfloor \frac{\mathcal{N}}{\mathcal{T}_{\mathcal{P}}} \rfloor$ sub-trajectories. $\mathcal{T}_{\mathcal{P}}$ is data-dependent and has no definite value. For example, $\mathcal{T}_{\mathcal{P}}$ can be set to 'a day' in traffic control applications since many vehicles have daily patterns, while annual animal migration patterns can be discovered by $\mathcal{T}_{\mathcal{P}}$ = 'a year'. For instance, see Figure 3, an object's trajectory is decomposed into daily sub-trajectories.

Essentially, a periodic pattern is a closed swarm discovered from $\lfloor \frac{\mathcal{N}}{\mathcal{T}_{\mathcal{P}}} \rfloor$ sub-trajectories. For instance, in Figure 3, we have 3 daily sub-trajectories and from them we extract the two following periodic patterns $\{c_1, c_2, c_3, c_4\}$ and $\{c_1, c_3, c_4\}$. The main difference in periodic pattern mining is the preprocessing data step while the definition is similar to that of a closed swarm. As we have provided the definition of a closed swarm, we will mainly focus on closed swarm mining below.

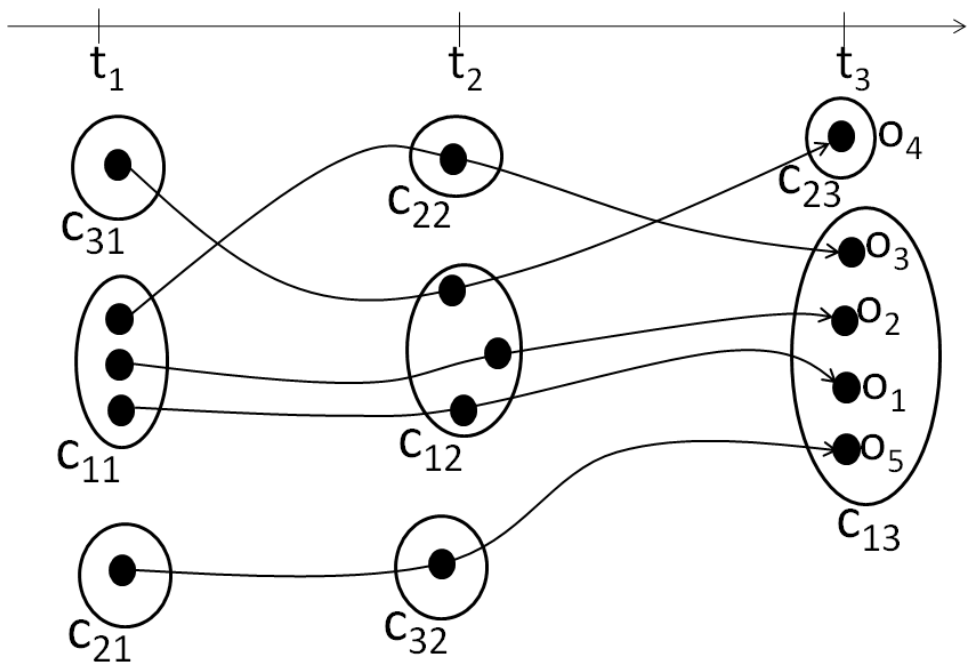

**Figure 4: An illustrative example.**

## 2.2 Related Work

As mentioned before, many approaches have been proposed to extract patterns. The interested reader may refer to [20, 28] where short descriptions of the most efficient or interesting patterns and approaches are proposed. For instance, Gudmundsson and van Kreveld [1], Vieira et al. [2] define a flock pattern, in which the same set of objects stay together in a circular region with a predefined radius, Kalnis et al. [4] propose the notion of *moving clusters*, while Jeung et al. [3] define a convoy pattern.

Jeung et al. [3] adopt the DBScan algorithm [5] to find candidate convoy patterns. The authors propose three algorithms that incorporate trajectory simplification techniques in the first step. The distance measurements are performed on trajectory segments of as opposed to point based distance measurements. Another problem is related to the trajectory representation. Some trajectories may have missing timestamps or are measured at different time intervals. Therefore, the density measurements cannot be applied between trajectories with different timestamps. To address the problem of missing timestamps, the authors proposed to interpolate the trajectories by creating virtual time points and by applying density measurements on trajectory segments. Additionally, the convoy is defined as a candidate when it has at least $k$ clusters during $k$ consecutive timestamps.

Recently, Zhenhui Li et al. [6] propose the concept of swarm and closed swarm and the *ObjectGrowth* algorithm to extract closed swarm patterns. The ObjectGrowth method is a depth-first-search framework based on the objectset search space (i.e., the collection of all subsets of $O_{DB}$). For the search space of $O_{DB}$, they perform depth-first search of all subsets of $O_{DB}$ through a pre-order tree traversal. Even though, the search space remains still huge for enumerating the objectsets in $O$ ($2^{|O_{DB}|}$). To speed up the search process, they propose two pruning rules. The first pruning rule, called *Apriori Pruning*, is used to stop traversal the subtree when we find further traversal that cannot satisfy $min_t$. The second pruning rule, called *Backward Pruning*, makes use of the closure property. It checks whether there is a superset of the current objectset, which has the same maximal corresponding timeset as that of the current one. If so, the traversal of the subtree under the current objectset is meaningless. After pruning the invalid candidates, the remaining ones may or may not be closed swarms. Then a *Forward Closure Checking* is used to determine whether a pattern is a closed swarm or not.

In [21], Hwang et al. propose two algorithms to mine group patterns, known as the *Apriori-like Group Pattern mining* algorithm and *Valid Group-Growth* algorithm. The former explores the Apriori property of valid group patterns and extends the Apriori algorithm [11] to mine valid group patterns. The latter is based on idea similar to the FP-growth algorithm [27].

Recently in [29], A. Calmeron proposes a frequent itemset-based approach for flock identification purposes.

Even if these approaches are very efficient they suffer the problem that they only extract a specific kind of pattern. When considering a dataset, it is quite difficult, for the decision maker, to know in advance the kind of pattern embedded in the data. Therefore proposing an approach able to automatically extract all these different kinds of patterns can be very useful and this is the problem we address in this paper and that will be developed in the next sections.

**Table 2: Cluster Matrix**

| $T_{DB}$ | | $t_1$ | | | $t_2$ | | | $t_3$ | |
|---|---|---|---|---|---|---|---|---|---|
| Clusters $C_{DB}$ | | $c_{11}$ | $c_{21}$ | $c_{31}$ | $c_{12}$ | $c_{22}$ | $c_{32}$ | $c_{13}$ | $c_{23}$ |
| $O_{DB}$ | $o_1$ | 1 | | | 1 | | | 1 | |
| | $o_2$ | 1 | | | 1 | | | 1 | |
| | $o_3$ | 1 | | | | 1 | | 1 | |
| | $o_4$ | | | 1 | 1 | | | | 1 |
| | $o_5$ | | 1 | | | | 1 | 1 | |

# 3. SPATIO-TEMPORAL PATTERNS IN ITEMSET CONTEXT

Extracting different kinds of patterns requires the use of several algorithms and to deal with this problem, we propose an unifying approach to extract and manage different kinds of patterns.

Basically, patterns are evolution of clusters over time. Therefore, to manage the evolution of clusters, we need to analyse the correlations between them. Furthermore, if clusters share some characteristics (e.g. share some objects), they could be a pattern. Consequently, if a cluster is considered as an item we will have a set of items (called itemset). The main problem essentially is to efficiently combine items (clusters) to find itemsets (a set of clusters) which share some characteristics or satisfy some properties to be considered a pattern. To describe cluster evolution, spatio-temporal data is presented as a cluster matrix from which patterns can be extracted.

Definition 5. *Cluster Matrix. Assume that we have a set of clusters* $C_{DB} = \{C_1, C_2, \ldots, C_n\}$ *where* $C_i = \{c_{i_1 t_i}, c_{i_2 t_i}, \ldots, c_{i_m t_i}\}$ *is a set of clusters at timestamps* $t_i$. *A cluster matrix is thus a matrix of size* $|O_{DB}| \times |C_{DB}|$. *Each row represents an object and each column represents a cluster. The value of the cluster matrix cell,* $(o_i, c_j)$ *is 1 (resp. empty) if* $o_i$ *is in (resp. is not in) cluster* $c_j$. *A cluster (or item)* $c_j$ *is a cluster formed after applying clustering techniques.*

For instance, the data from illustrative example (Figure 4) is presented in a cluster matrix in Table 2. Object $o_1$ belongs to the cluster $c_{11}$ at timestamp $t_1$. For clarity reasons in the following, $c_{ij}$ represents the cluster $c_i$ at time $t_j$. Therefore, the matrix cell ($o_1$-$c_{11}$) is 1, meanwhile the matrix cell ($o_4$-$c_{11}$) is empty because object $o_4$ does not belong to cluster $c_{11}$.

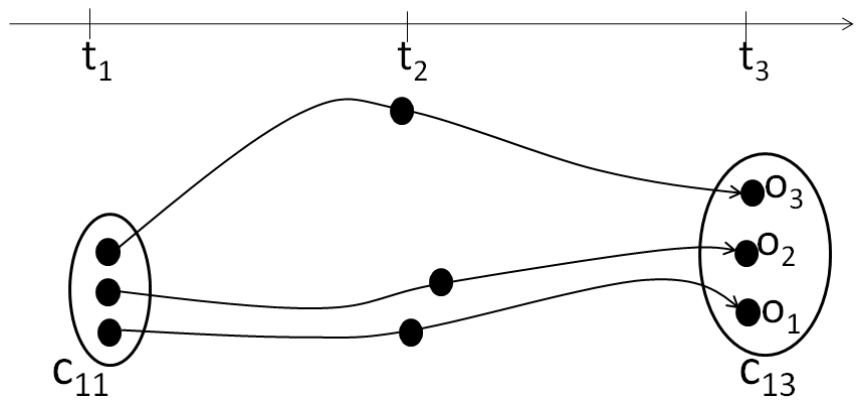

**Figure 5: A swarm from our example.**

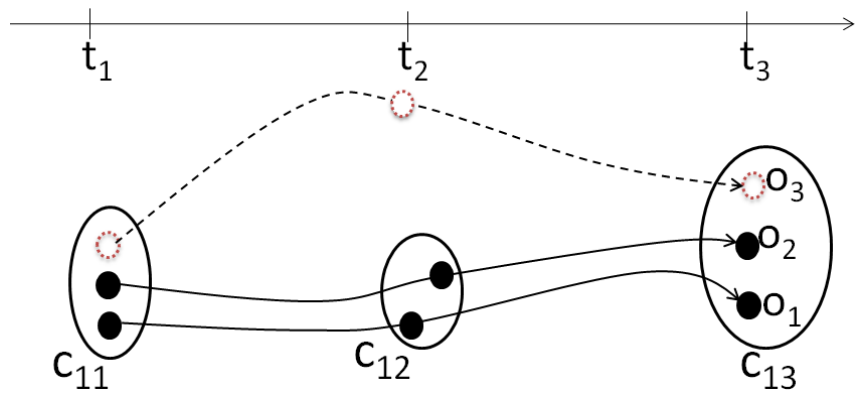

**Figure 6: A convoy from our example.**

By presenting data in a cluster matrix, each object acts as a transaction while each cluster $c_j$ stands for an item. Additionally, an itemset can be formed as $\Upsilon = \{c_{t_{a_1}}, c_{t_{a_2}}, \dots, c_{t_{a_p}}\}$ with life time $T_\Upsilon = \{t_{a_1}, t_{a_2}, \dots, t_{a_p}\}$ where $t_{a_1} < t_{a_2} < \dots < t_{a_p}$, $\forall a_i : t_{a_i} \in T_{DB}, c_{t_{a_i}} \in C_{a_i}$. The support of the itemset $\Upsilon$, denoted $\sigma(\Upsilon)$, is the number of common objects in every items belonging to $\Upsilon$, $O(\Upsilon) = \bigcap_{i=1}^{p} c_{t_{a_i}}$. Additionally, the length of $\Upsilon$, denoted $|\Upsilon|$, is the number of items or timestamps ($= |T_\Upsilon|$).

For instance, in Table 2, for a support value of 2 we have: $\Upsilon = \{c_{11}, c_{12}\}$ veryfying $\sigma(\Upsilon) = 2$. Every items (resp. clusters) of $\Upsilon, c_{11}$ and $c_{12}$, are in the transactions (resp. objects) $o_1, o_2$. The length of $|\Upsilon|$ is the number of items ($= 2$).

Naturally, the number of clusters can be large; however, the maximum length of itemsets is $|T_{DB}|$. Because of the density-based clustering algorithm used, clusters at the same timestamp cannot be in the same itemsets.

Now, we will define some useful properties to extract the patterns presented in Section 2 from frequent itemsets as follows:

Property 1. *Swarm. Given a frequent itemset* $\Upsilon = \{c_{t_{a_1}}, c_{t_{a_2}}, \dots, c_{t_{a_p}}\}$. $(O(\Upsilon), T_\Upsilon)$ *is a swarm if, and only if:*

$$\begin{cases} (1) : \sigma(\Upsilon) \geq \varepsilon \\ (2) : |\Upsilon| \geq min_t \end{cases} \quad (5)$$

Proof. After construction, we have $\sigma(\Upsilon) \geq \varepsilon$ and $\sigma(\Upsilon) = |O(\Upsilon)|$ then $|O(\Upsilon)| \geq \varepsilon$. Additionally, as $|\Upsilon| \geq min_t$ and $|\Upsilon| = |T_\Upsilon|$ then $|T_\Upsilon| \geq min_t$. Furthermore, $\forall t_{a_j} \in T_\Upsilon, O(\Upsilon) \subseteq c_{t_{a_j}}$, means that at every timestamp we have a cluster containing all objects in $O(\Upsilon)$. Consequently, $(O(\Upsilon), T_\Upsilon)$ is a swarm because it satisfies all the requirements of the *Definition 1*. □

For instance, in Figure 5, for the frequent itemset $\Upsilon = \{c_{11}, c_{13}\}$ we have $(O(\Upsilon) = \{o_1, o_2, o_3\}, T_\Upsilon = \{t_1, t_3\})$ which is a swarm with support threshold $\varepsilon = 2$ and $min_t = 2$. We can notice that $\sigma(\Upsilon) = 3 > \varepsilon$ and $|\Upsilon| = 2 \geq min_t$.

Essentially, a closed swarm is a swarm which satisfies the *object-closed* and *time-closed* conditions therefore closed-swarm property is as follows:

Property 2. *Closed Swarm. Given a frequent itemset* $\Upsilon = \{c_{t_{a_1}}, c_{t_{a_2}}, \dots, c_{t_{a_p}}\}$. $(O(\Upsilon), T_\Upsilon)$ *is a closed swarm if and only if:*

$$\begin{cases} (1) : (O(\Upsilon), T_\Upsilon) \text{ is a swarm.} \\ (2) : \nexists \Upsilon' \text{ s.t } O(\Upsilon) \subset O(\Upsilon'), T_{\Upsilon'} = T_\Upsilon \text{ and} \\ (O(\Upsilon'), T_\Upsilon) \text{ is a swarm.} \\ (3) : \nexists \Upsilon' \text{ s.t. } O(\Upsilon') = O(\Upsilon), T_\Upsilon \subset T_{\Upsilon'} \text{ and} \\ (O(\Upsilon), T_{\Upsilon'}) \text{ is a swarm.} \end{cases} \quad (6)$$

Proof. After construction, we obtain $(O(\Upsilon), T_\Upsilon)$ which is a swarm. Additionally, if $\nexists \Upsilon'$ s.t $O(\Upsilon) \subset O(\Upsilon'), T_{\Upsilon'} = T_\Upsilon$ and $(O(\Upsilon'), T_\Upsilon)$ is a swarm then $(O(\Upsilon), T_\Upsilon)$ cannot be enlarged in terms of objects. Therefore, it satisfies the *object-closed* condition. Furthermore, if $\nexists \Upsilon'$ s.t. $O(\Upsilon') = O(\Upsilon), T_\Upsilon \subset T_{\Upsilon'}$ and $(O(\Upsilon), T_{\Upsilon'})$ is a swarm then $(O(\Upsilon), T_\Upsilon)$ cannot be enlarged in terms of lifetime. Therefore, it satisfies the *time-closed* condition. Consequently, $(O(\Upsilon), T_\Upsilon)$ is a swarm and it satisfies *object-closed* and *time-closed* conditions and therefore $(O(\Upsilon), T_\Upsilon)$ is a closed swarm according to the *Definition 6*. □

According to the convoy *Definition 3*, a convoy is a swarm which satisfies the consecutiveness condition. Therefore, for an itemset, we can extract a convoy if the following property holds:

Property 3. *Convoy. Given a frequent itemset* $\Upsilon = \{c_{t_{a_1}}, c_{t_{a_2}}, \dots, c_{t_{a_p}}\}$. $(O(\Upsilon), T_\Upsilon)$ *is a convoy if and only if:*

$$\begin{cases} (1) : (O(\Upsilon), T_\Upsilon) \text{ is a swarm.} \\ (2) : \forall j, 1 \leq j < p : t_{a_j}, t_{a_{j+1}} \text{ are consecutive.} \end{cases} \quad (7)$$

Proof. After construction, we obtain $(O(\Upsilon), T_\Upsilon)$ which is a swarm. Additionally, if $\Upsilon$ satisfies the condition (2), it means that the $\Upsilon$'s lifetime is consecutive. Consequently, $(O(\Upsilon), T_\Upsilon)$ is a convoy according to the *Definition 3*. □

For instance, see Table 2 and Figure 6, for the frequent itemset $\Upsilon = \{c_{11}, c_{12}, c_{13}\}$ we have $(O(\Upsilon) = \{o_1, o_2\}, T_\Upsilon = \{t_1, t_2, t_3\})$ is a convoy with support threshold $\varepsilon = 2$ and $min_t = 2$. Note that $o_3$ is not in the convoy.

Please remember that group pattern is a set of disjointed convoys which share the same objects, but in different time intervals. Therefore, the group pattern property is as follows:

Property 4. *Group Pattern. Given a frequent itemset* $\Upsilon = \{c_{t_{a_1}}, c_{t_{a_2}}, \dots, c_{t_{a_p}}\}$*, a mininum weight* $min_{wei}$*, a minimum number of convoys* $min_c$*, a set of consecutive time segments* $T_\mathcal{S} = \{s_1, s_2, \dots, s_n\}$. $(O(\Upsilon), T_\mathcal{S})$ *is a group pattern if and only if:*

$$\begin{cases} (1) : |T_\mathcal{S}| \geq min_c. \\ (2) : \forall s_i, s_i \subseteq T_\Upsilon, |s_i| \geq min_t. \\ (3) : \bigcap_{i=1}^{n} s_i = \emptyset, \bigcap_{i=1}^{n} O(s_i) = O(\Upsilon). \\ (4) : \forall s \notin T_\mathcal{S}, s \text{ is a convoy, } O(\Upsilon) \nsubseteq O(s). \\ (5) : \frac{\sum_{i=1}^{n} |s_i|}{|T|} \geq min_{wei}. \end{cases} \quad (8)$$

Proof. If $|T_\mathcal{S}| \geq min_c$ then we know that at least $min_c$ consecutive time intervals $s_i$ in $T_\mathcal{S}$. Furthermore, if $\forall s_i, s_i \subseteq T_\Upsilon$ then we have $O(\Upsilon) \subseteq O(s_i)$. Additionally, if $|s_i| \geq min_t$ then $(O(\Upsilon), s_i)$ is a convoy (*Definition 3*). Now, $T_\mathcal{S}$ actually is a set of convoys of $O(\Upsilon)$ and if $\bigcap_{i=1}^{n} s_i = \emptyset$ then $T_\mathcal{S}$ is a set of disjointed convoys. A little bit further, if $\forall s \notin T_{\mathcal{S}}, s$ is a convoy and $O(\Upsilon) \not\subseteq O(s)$ then $\nexists T_{\mathcal{S}'}$ s.t. $T_{\mathcal{S}} \subset T_{\mathcal{S}'}$ and $\bigcap_{i=1}^{|T_{\mathcal{S}'}|} O(s_i) = O(\Upsilon)$. Therefore, $(O(\Upsilon), T_{\mathcal{S}})$ cannot be enlarged in terms of *number of convoys*. Similarly, if $\bigcap_{i=1}^{n} O(s_i) = O(\Upsilon)$ then $(O(\Upsilon), T_{\mathcal{S}})$ cannot be enlarged in terms of *objects*. Consequently, $(O(\Upsilon), T_{\mathcal{S}})$ is a closed swarm of disjointed convoys because $|O(\Upsilon)| \geq \varepsilon, |T_{\mathcal{S}}| \geq min_c$ and $(O(\Upsilon), T_{\mathcal{S}})$ cannot be enlarged (*Definition 6*). Finally, if $(O(\Upsilon), T_{\mathcal{S}})$ satisfies condition (5) then it is a valid group pattern due to *Definition 4*. □

**Table 3: Periodic Cluster Matrix**

| $T_{DB}$ | | $t_1$ | $t_2$ | | $t_3$ |
|---|---|---|---|---|---|
| Clusters $C_{DB}$ | | $c_{11}$ | $c_{12}$ | $c_{22}$ | $c_{13}$ |
| $ST_{DB}$ | $st_1$ | 1 | 1 | | 1 |
| | $st_2$ | 1 | 1 | | 1 |
| | $st_3$ | 1 | | 1 | 1 |

As mentioned before, the main difference in periodic pattern mining is the input data while the definition is similar to that of closed swarm. The cluster matrix which is used for periodic mining can be defined as follows:

Definition 6. *Periodic Cluster Matrix (PCM). Periodic cluster matrix is a cluster matrix with some differences as follows: 1) Object o is a sub-trajectory st, 2) $ST_{DB}$ is a set of all sub-trajectories in dataset.*

For instance, see Table 3, an object's trajectory is decomposed into 3 sub-trajectories and from them a periodic cluster matrix can be generated by applying clustering technique. Assume that we can extract a frequent itemsets $\Upsilon = \{c_{t_{a_1}}, c_{t_{a_2}}, \ldots, c_{t_{a_p}}\}$ from periodic cluster matrix, the periodic can be defined as follows:

Property 5. *Periodic Pattern. Given a frequent itemset $\Upsilon = \{c_{t_{a_1}}, c_{t_{a_2}}, \ldots, c_{t_{a_p}}\}$, a mininum weight $min_{wei}$which is extracted from periodic cluster matrix. $(ST(\Upsilon), (T)_\Upsilon)$ is a periodic pattern if and only if $(ST(\Upsilon), (T)_\Upsilon)$ is a closed swarm. Note that $ST(\Upsilon) = \bigcap_{i=1}^{p} c_{t_{a_i}}$*

Above, we presented some useful properties to extract spatio-temporal patterns from itemsets. Now we will focus on the fact that from an itemset mining algorithm we are able to extract the set of all spatio-temporal patterns. We thus start the proof process by analyzing the swarm extracting problem. This first lemma shows that from a set of frequent itemsets we are able to extract all the swarms embedded in the database.

Lemma 1. *Let $FI = \{\Upsilon_1, \Upsilon_2, \ldots, \Upsilon_l\}$ be the frequent itemsets being mined from the cluster matrix with minsup = $\varepsilon$. All swarms $(O, T)$ can be extracted from $FI$.*

Proof. Let us assume that $(O, T)$ is a swarm. Note, $T = \{t_{a_1}, t_{a_2}, \ldots, t_{a_m}\}$. According to the *Definition 1* we know that $|O| \geq \varepsilon$. If $(O, T)$ is a swarm then $\forall t_{a_i} \in T, \exists c_{t_{a_i}}$ s.t. $O \subseteq c_{t_{a_i}}$ therefore $\bigcap_{i=1}^{m} c_{t_{a_i}} = O$. Additionally, we know that $\forall c_{t_{a_i}}$, $c_{t_{a_i}}$ is an item so $\exists \Upsilon = \bigcup_{i=1}^{m} c_{t_{a_i}}$ is an itemset and $O(\Upsilon) = \bigcap_{i=1}^{m} c_{t_{a_i}} = O, T_\Upsilon = \bigcup_{i=1}^{m} t_{ai} = T$. Therefore, $(O(\Upsilon), T_\Upsilon)$ is a swarm. So, $(O, T)$ is extracted from $\Upsilon$. Furthermore, $\sigma(\Upsilon) = |O(\Upsilon)| = |O| \geq \varepsilon$ then $\Upsilon$ is a frequent itemset and $\Upsilon \in FI$. Finally, $\forall (O, T)$ s.t. if $(O, T)$ is a swarm then $\exists \Upsilon$ s.t. $\Upsilon \in FI$ and $(O, T)$ can be extracted from $\Upsilon$, we can conlude $\forall$ swarm $(O, T)$, it can be mined from $FI$. □

We can consider that by adding constraints such as "consecutive lifetime", "time-closed", "object-closed", "integrity proportion" to swarms, we can retrieve convoys, closed swarms and moving clusters. Therefore, if $Swarm, CSwarm, Convoy, MCluster$ respectively contain all swarms, closed-swarms, convoys and moving clusters then we have: $CSwarm \subseteq Swarm$, $Convoy \subseteq Swarm$ and $MCluster \subseteq Swarm$. By applying *Lemma 1*, we retrieve all swarms from frequent itemsets. Since, a set of closed swarms, a set of convoys and a set of moving clusters are subsets of swarms and they can therefore be completely extracted from frequent itemsets. Additionally, all periodic patterns also can be extracted because they are similar to closed swarms. Now, we will consider group patterns and we show that all of them can be directly extracted from the set of all frequent itemsets.

Lemma 2. *Given $FI = \{\Upsilon_1, \Upsilon_2, \ldots, \Upsilon_l\}$ contains all frequent itemsets mined from cluster matrix with minsup = $\varepsilon$. All group patterns $(O, T_{\mathcal{S}})$ can be extracted from $FI$.*

Proof. $\forall (O, T_{\mathcal{S}})$ is a valid group pattern, we have $\exists T_{\mathcal{S}} = \{s_1, s_2, \ldots, s_n\}$ and $T_{\mathcal{S}}$ is a set of disjointed convoys of $O$. Therefore, $(O, T_{s_i})$ is a convoy and $\forall s_i \in T_{\mathcal{S}}, \forall t \in T_{s_i}, \exists c_t$ s.t. $O \subseteq c_t$. Let us assume $C_{s_i}$ is a set of clusters corresponding to $s_i$, we know that $\exists \Upsilon$, $\Upsilon$ is an itemset, $\Upsilon = \bigcup_{i=1}^{n} C_{s_i}$ and $O(\Upsilon) = \bigcap_{i=1}^{n} O(C_{s_i}) = O$. Additionally, $(O, T_{\mathcal{S}})$ is a valid group pattern; therefore, $|O| \geq \varepsilon$ so $|O(\Upsilon)| \geq \varepsilon$. Consequently, $\Upsilon$ is a frequent itemset and $\Upsilon \in FI$ because $\Upsilon$ is an itemset and $\sigma(\Upsilon) = |O(\Upsilon)| \geq \varepsilon$. Consequently, $\forall (O, T_{\mathcal{S}}), \exists \Upsilon \in FI$ s.t. $(O, T_{\mathcal{S}})$ can be extracted from $\Upsilon$ and therefore all group patterns can be extracted from $FI$. □

As we have shown that patterns such as swarms, closed swarms, convoys, group patterns can be similarly mapped into frequent itemset context. However, mining all frequent itemsets is cost prohibitive in some cases. Moreover, the set of frequent closed itemsets has been proved to be a condensed collection of frequent itemsets, i.e., both a concise and lossless represention of a collection of frequent itemsets [8, 9, 10, 24, 26, 30]. They are concise since a collection of frequent closed itemsets is orders of magnitude smaller than the corresponding collection of frequents. This allows the use of very low minimum support thresholds. Moreover, they are lossless, because it is possible to derive the identity and the support of every frequent itemsets in the collection from them. Therefore, we only need to extract frequent closed itemsets and then to scan them with properties to obtain the corresponding spatio-temporal patterns instead of having to mine all frequent itemsets.

# 4. FREQUENT CLOSED ITEMSET-BASED SPATIO-TEMPORAL PATTERN MINING ALGORITHM

Recently, patterns have been redefined in the itemset context. In this section, we propose two approaches i.e., *GeT_Move* and *Incremental GeT_Move*, to efficiently extract patterns. The global process is illustrated in Figure 7.

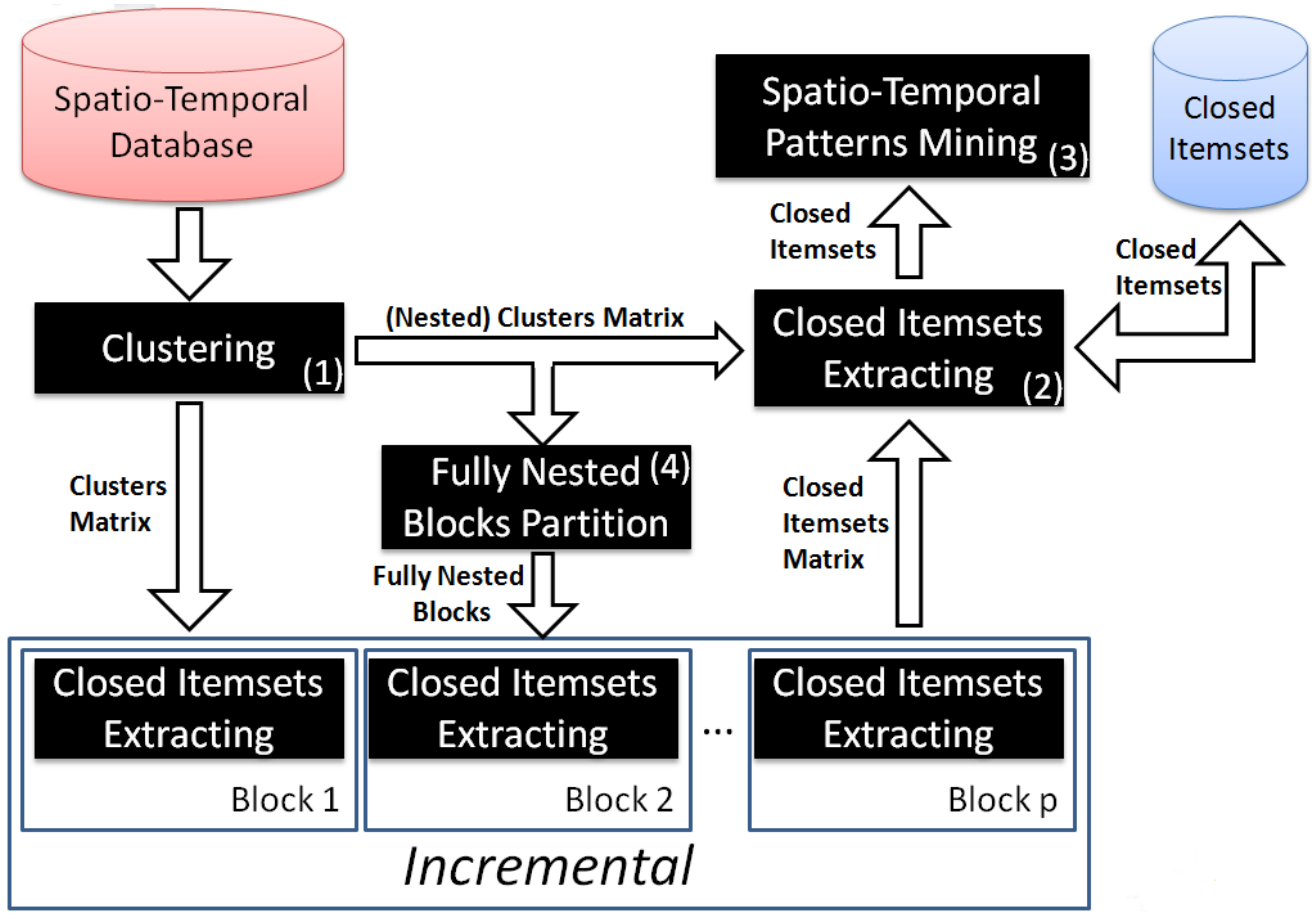

**Figure 7: The main process.**

In the first step, a clustering approach (Figure 7-(1)) is applied at each timestamp to group objects into different clusters. For each timestamp $t_a$, we have a set of clusters $C_a = \{c_{1t_a}, c_{2t_a}, \ldots, c_{mt_a}\}$, with $1 \leq k \leq m, c_{kt_a} \subseteq O_{DB}$. Spatio-temporal data can thus be converted to a cluster matrix $CM$ (Table 2).

## 4.1 GeT_Move

After generating the cluster matrix $CM$, a frequent closed itemset mining algorithm is applied on $CM$ to extract all the frequent closed itemsets. By scanning frequent closed itemsets and checking properties, we can obtain the patterns.

In this paper, we apply the LCM algorithm [26, 30] to extract frequent closed itemsets as it is known to be a very efficient algorithm. The key feature of the LCM algorithm is that after discovering a frequent closed itemset $X$, it generates a new generator $X[i]$ by extending $X$ with a frequent item $i, i \notin X$. Using a total order relation on frequent items, LCM verifies if $X[i]$ violates this order by performing tests using only the *tidset*[2] of $X$, called $\mathcal{T}(X)$, and those of the frequent items $i$. If $X[i]$ is not discarded, then $X[i]$ is an order preserving generator of a new frequent closed itemset. Then, its closure is computed using the previously mentioned tidsets.

In this process, we discard some useless candidate itemsets. In spatio-temporal patterns, items (resp. clusters) must belong to different timestamps and therefore items (resp. clusters) which form a FCI must be in different timestamps. In contrast, we are not able to extract patterns by combining items in the same timestamp. Consequently, FCIs which include more than 1 item in the same timestamp will be discarded.

Thanks to the above characteristic, we now have the maximum length of the frequent closed itemsets which is the number of timestamps $|T_{DB}|$. Additionally, the LCM search space only depends on the number of objects (transactions) $|O_{DB}|$ and the maximum length of itemsets $|T_{DB}|$. Consequently, by using LCM and by applying the property, *GeT_Move* is not affected by the number of clusters and therefore the computing time can be greatly reduced.

The pseudo code of *GeT_Move* is described in *Algorithm 1*. The core of *GeT_Move* algorithm is based on the LCM algorithm which has been slightly modified by adding the pruning rule and by extracting patterns from FCIs. The initial value of FCI $X$ is empty and then we start by putting item $i$ into $X$ (lines 2-3). By adding $i$ into $X$, we have $X[i]$ and if $X[i]$ is a FCI then $X[i]$ is used as a generator of a new FCI, call LCM_Iter$(X, \mathcal{T}(X), i(X))$ (lines 4-5). In LCM_Iter, we first check properties of Section 3 (line 8) for FCI $X$. Next, for each transaction $t \in \mathcal{T}(X)$, we add all items $j$, which are larger than $i(X)$ and satisfy the pruning rule, into occurrence sets $\mathcal{J}[j]$ (lines 9-11). Next, for each $j \in \mathcal{J}[j]$, we check to see if $\mathcal{J}[j]$ is a FCI, and if so, then we recall LCM_Iter with the new generator (lines 12-14). After terminating the call for $\mathcal{J}[j]$, the memory for $\mathcal{J}[j]$ is released for the future use in $\mathcal{J}[k]$ for $k < j$ (lines 15).

Regarding to the PatternMining sub-function (lines 16-37), the algorithm basically checks properties of the itemset $X$ to extract spatio-temporal patterns. If $X$ satisfies the $min_t$ condition then $X$ is a closed swarm (lines 18-19). After that, we check the consecutive time constraint for convoy and moving cluster (lines 21-22) and then if the convoy satisfies $min_t$ condition and correctness in terms of objects containing (line 31), output convoy (line 32). Next, we put convoy into group pattern *gPattern* (line 33) and then output group pattern if it satisfies the $min_c$ condition and $min_{wei}$ condition at the end of scanning $X$ (line 37). Regarding to moving cluster $mc$, we check the integrity at each pair of consecutive timestamps (line 24). If $mc$ satisfies the condition then the previous item $x_k$ will be merged into $mc$ (line 25). If not, we check the $min_t$ condition for $mc \cup x_k$ and if it is satisfied then we output $mc \cup x_k$ as a moving cluster.

## 4.2 Incremental GeT_Move

Naturally, in real world applications (cars, animal migration), the objects tend to move together in short interval meanwhile their movements can be different in long interval. Therefore, the number of items (clusters) can be large and the length of FCIs can be long. For instance, let us consider the Figure 8a, objects $\{o_1, o_2, o_3, o_4\}$ move together during first 100 timestamps and after that $o_1, o_2$ stay together while $o_3, o_4$ move together in another direction. The problem here is that if we apply *GeT_Move* on the whole dataset, the extraction of the itemsets can be very time consuming.

To deal with this issue, we propose the *Incremental GeT_Move* algorithm. The main idea is to split the trajectories (resp. cluster matrix CM) into short intervals, called blocks. By applying frequent closed itemset mining on each short interval, the data can then be compressed into local frequent closed itemsets. Additionally, the length of itemsets and the number of items can be greatly reduced.

For instance, see Figure 8, if we consider $[t_1, t_{100}]$ as a block and $[t_{101}, t_{200}]$ as another block, the maximum length of itemsets in both blocks is 100 (insteads of 200). Additionally, the original data can be greatly compressed (e.g. Figure 8b) and only 3 items remain: $ci_{11}, ci_{12}, ci_{22}$. Consequently, the process is much improved.

Definition 7. *Block. Given a set of timestamps $T_{DB} = \{t_1, t_2, \ldots, t_n\}$, a cluster matrix $CM$. $CM$ is vertically split into equivalent (in terms of intervals) smaller cluster matrices and each of them is a block $b$. Assume $T_b$ is a set of timestamps of block $b$, $T_b = \{t_1, t_2, \ldots, t_k\}$, thus we have $|T_b| = k \leq |T_{DB}|$.*

[2]Called *tidlists* in [24, 10] and *denotations* in [26, 30].

```
Algorithm 1: GeT_Move
Input : Occurrence sets 𝒥, int ε, int min_t, set of
        items C_DB, double θ, int min_c, double min_wei
begin
  | X := I(𝒯(∅)); //The root
  | for i := 1 to |C_DB| do
  |   | if |𝒯(X[i])| ≥ ε and |X[i]| is closed then
  |   |   | LCM_Iter(X[i], 𝒯(X[i]), i);
LCM_Iter(X, 𝒯(X), i(X))
begin
  | PatternMining(X, min_t); /*X is a pattern?*/
  | foreach transaction t ∈ 𝒯(X) do
  |   | foreach j ∈ t, j > i(X), j.time ∉ time(X) do
  |   |   | insert j to 𝒥[j];
  | foreach j ∈ 𝒥[j] in the decreasing order do
  |   | if |𝒯(𝒥[j])| ≥ ε and 𝒥[j] is closed then
  |   |   | LCM_Iter(𝒥[j], 𝒯(𝒥[j]), j);
  |   | Delete 𝒥[j];
PatternMining(X, min_t)
begin
  | if |X| ≥ min_t then
  |   | output X; /*Closed Swarm*/
  |   | gPattern := ∅; convoy := ∅; mc := ∅;
  |   | for k := 1 to |X − 1| do
  |   |   | if x_k.time = x_(k+1).time − 1 then
  |   |   |   | convoy := convoy ∪ x_k;
  |   |   |   | if |𝒯(x_k) ∩ 𝒯(x_{k+1})| / |𝒯(x_k) ∪ 𝒯(x_{k+1})| ≥ θ then
  |   |   |   |   | mc := mc ∪ x_k;
  |   |   |   | else
  |   |   |   |   | if |mc ∪ x_k| ≥ min_t then
  |   |   |   |   |   | output mc ∪ x_k;
  |   |   |   |   |   | /*MovingCluster*/
  |   |   |   |   | mc := ∅;
  |   |   | else
  |   |   |   | if |convoy ∪ x_k| ≥ min_t and
  |   |   |   | |𝒯(convoy ∪ x_k)| = |𝒯(X)| then
  |   |   |   |   | output convoy ∪ x_k; /*Convoy*/
  |   |   |   |   | gPattern := gPattern ∪ (convoy ∪ x_k);
  |   |   |   | if |mc ∪ x_k| ≥ min_t then
  |   |   |   |   | output mc ∪ x_k; /*MovingCluster*/
  |   |   |   | convoy := ∅; mc := ∅;
  |   | if |gPattern| ≥ min_c and
  |   | size(gPattern) / |T_DB| ≥ min_wei then
  |   |   | output gPattern; /*Group Pattern*/
Where: X is itemset, X[i] := X ∪ i, i(X) is the last
item of X, 𝒯(X) is list of tractions that X belongs to,
𝒥[j] := 𝒯(X[j]), j.time is time index of item j,
time(X) is a set of time indexes of X, |𝒯(convoy)| is
the number of transactions that the convoy belongs to,
|gPattern| and size(gPattern) respectively are the
number of convoys and the total length of the convoys
in gPattern.
```

Assume that we obtain a set of blocks $B = \{b_1, b_2, \ldots, b_p\}$ with $|T_{b_1}| = |T_{b_2}| = \ldots = |T_{b_p}|, \bigcup_{i=1}^{p} b_i = CM$ and $\bigcap_{i=1}^{p} b_i = \emptyset$. Given a set of frequent closed itemset collections $CI = \{CI_1, CI_2, \ldots, CI_p\}$ where $CI_i$ is mined from block $b_i$. $CI$ is presented as a *closed itemset matrix* which is formed by horizontally connecting all local frequent closed itemsets: $CIM = \bigcup_{i=1}^{p} CI_i$.

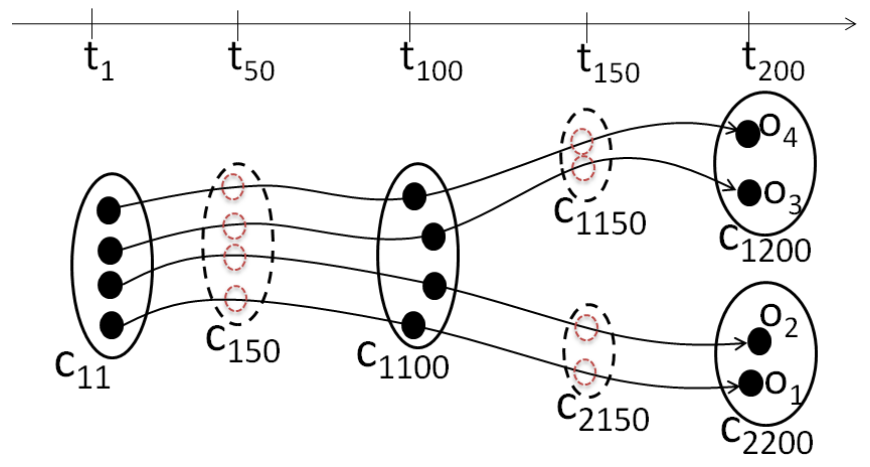

(a) The entire dataset.

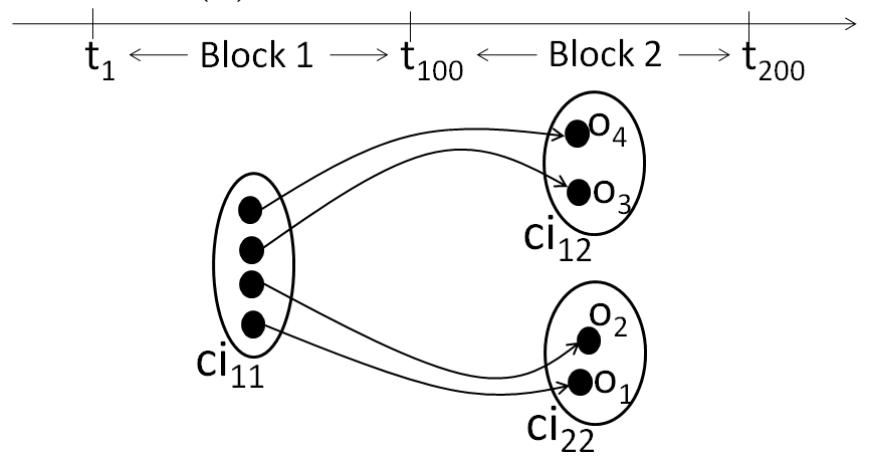

(b) Data after applying frequent closed itemsets mining on Blocks.

**Figure 8: A case study example. (b)-$ci_{11}$ (resp. $ci_{12}, ci_{22}$) is a frequent closed itemset extracted from block 1 (resp. block 2).**

**Table 4: Closed Itemset Matrix**

| Block $B$ | | $b_1$ | $b_2$ | |
|---|---|---|---|---|
| Frequent Closed Itemsets $CI$ | | $ci_{11}$ | $ci_{12}$ | $ci_{22}$ |
| $O_{DB}$ | $o_1$ | 1 | | 1 |
| | $o_2$ | 1 | | 1 |
| | $o_3$ | 1 | 1 | |
| | $o_4$ | 1 | 1 | |

Definition 8. *Closed Itemset Matrix (CIM). Closed itemset matrix is a cluster matrix with some differences as follows: 1) Timestamp $t$ now becomes a block $b$. 2) Item $c$ is a frequent closed itemset $ci$.*

For instance, see Table 4, we have two sets of frequent closed itemsets $CI_1 = \{ci_{11}\}, CI_2 = \{ci_{12}, ci_{22}\}$ which are respectively extracted from blocks $b_1, b_2$. Closed itemset matrix $CIM = CI_1 \cup CI_2$ means that $CIM$ is created by horizontally connecting $CI_1$ and $CI_2$. Consequently, we have $CIM$ as in Table 4.

We have already provided *blocks* to compress original data. Now, by applying frequent closed itemset mining on closed itemset matrix $CIM$, we are able to retrieve all frequent closed itemsets from corresponding data. Note that items (in $CIM$) which are in the same block cannot be in the same frequent closed itemsets.

Lemma 3. *Given a cluster matrix $CM$ which is vertically split into a set of blocks $B = \{b_1, b_2, \ldots, b_p\}$ so that $\forall \Upsilon, \Upsilon$ is a frequent closed itemset and $\Upsilon$ is extracted from $CM$ then $\Upsilon$ can be extracted from closed itemset matrix $CIM$.*

Proof. Let us assume that $\forall b_i, \exists I_i$ is a set of items belonging to $b_i$ and therefore we have $\bigcap_{i=1}^{|B|} I_i = \emptyset$. If $\forall \Upsilon, \Upsilon$ is a FCI extracted from $CM$ then $\Upsilon$ is formed as $\Upsilon = \{\gamma_1, \gamma_2, \ldots, \gamma_p\}$ where $\gamma_i$ is a set of items s.t. $\gamma_i \subseteq I_i$. Additionally, $\Upsilon$ is a FCI and $O(\Upsilon) = \bigcap_{i=1}^{p} O(\gamma_i)$ then $\forall O(\gamma_i), O(\Upsilon) \subseteq O(\gamma_i)$. Furthermore, we have $|O(\Upsilon)| \geq \varepsilon$; therefore, $|O(\gamma_i)| \geq \varepsilon$ so $\gamma_i$ is a frequent itemset. Assume that $\exists \gamma_i, \gamma_i \notin CI_i$ then $\exists \Psi, \Psi \in CI_i$ s.t. $\gamma_i \subseteq \Psi$ and $\sigma(\gamma_i) = \sigma(\Psi), O(\gamma_i) = O(\Psi)$. Note that $\Psi$, $\gamma_i$ are from $b_i$. Remember that $O(\Upsilon) = O(\gamma_1) \cap O(\gamma_2) \cap \ldots \cap O(\gamma_i) \cap \ldots \cap O(\gamma_p)$ then we have: $\exists \Upsilon'$ s.t. $O(\Upsilon') = O(\gamma_1) \cap O(\gamma_2) \cap \ldots \cap O(\Psi) \cap \ldots \cap O(\gamma_p)$. Therefore, $O(\Upsilon') = O(\Upsilon)$ and $\sigma(\Upsilon') = \sigma(\Upsilon)$. Additionally, we know that $\gamma_i \subseteq \Psi$ so $\Upsilon \subseteq \Upsilon'$. Consequently, we obtain $\Upsilon \subseteq \Upsilon'$ and $\sigma(\Upsilon) = \sigma(\Upsilon')$. Therefore, $\Upsilon$ is not a FCI. That violates the assumption and therefore we have: if $\exists \gamma_i, \gamma_i \notin CI_i$ therefore $\Upsilon$ is not a FCI. Finally, we can conclude that $\forall \Upsilon, \Upsilon = \{\gamma_1, \gamma_2, \ldots, \gamma_p\}$ is a FCI extracted from $CM$, $\forall \gamma_i \in \Upsilon$, $\gamma_i$ must belong to $CI_i$ and $\gamma_i$ is an item in closed itemset matrix $CIM$. Therefore, $\Upsilon$ can be retrieved by applying FCI mining on $CIM$. □

**Algorithm 2: Incremental GeT_Move**

**Input** : Occurrence sets $K$, int $\varepsilon$, int $min_t$, double $\theta$, set of Occurrence sets (blocks) $B$, int $min_c$, double $min_{wei}$

1 **begin**
2 $K := \emptyset; CI := \phi$; int $item_total := 0$;
3 **foreach** $b \in B$ **do**
4 LCM($b, \varepsilon, I_b$);
5 GeT_Move($K, \varepsilon, min_t, CI, \theta, min_c, min_{wei}$);
6 **LCM(Occurrence sets $\mathcal{J}$, int $\sigma_0$, set of items $C$)**
7 **begin**
8 $X := I(\mathcal{T}(\emptyset))$; //The root
9 **for** $i := 1$ *to* $|C|$ **do**
10 **if** $|\mathcal{T}(X[i])| \geq \varepsilon$ *and* $|X[i]|$ *is closed* **then**
11 LCM_Iter($X[i], \mathcal{T}(X[i]), i$);
12 **LCM_Iter($X, \mathcal{T}(X), i(X)$)**
13 **begin**
14 **Update**($K, X, \mathcal{T}(X), item_total++$);
15 **foreach** *transaction* $t \in \mathcal{T}(X)$ **do**
16 **foreach** $j \in t, j > i(X), j.time \notin time(X)$ **do**
17 insert $j$ to $\mathcal{J}[j]$;
18 **foreach** $j, \mathcal{J}[j] \neq \phi$ *in the decreasing order* **do**
19 **if** $|\mathcal{T}(\mathcal{J}[j])| \geq \varepsilon$ *and* $\mathcal{J}[j]$ *is closed* **then**
20 LCM_Iter($\mathcal{J}[j], \mathcal{T}(\mathcal{J}[j]), j$);
21 **Delete** $\mathcal{J}[j]$;
22 **Update($K, X, \mathcal{T}(X), item_total$)**
23 **begin**
24 **foreach** $t \in \mathcal{T}(X)$ **do**
25 insert $item_total$ into $K[t]$;
26 $CI := CI \cup item_total$;

By applying *Lemma 3*, we can obtain all the FCIs and from the itemsets, patterns can be extracted. Note that the Incremental GeT_Move does not depend on the length restriction $min_t$. The reason is that $min_t$ is only used in Spatio-Temporal Patterns Mining step. Whatever $min_t$ ($min_t \geq$ block size or $min_t \leq$ block size), Incremental GeT_Move can extract all the FCIs and therefore the final results are the same.

The pseudo code of *Incremental GeT_Move* is described in *Algorithm 2*. The main difference between the code of *Incremental GeT_Move* and *GeT_Move* is the *Update* sub-function. In this function, we, step by step, generate the closed itemsets matrix from blocks (line 14 and lines 22-26). Next, we apply *GeT_Move* to extract patterns (line 5).

## 4.3 Toward A Parameter Free Incremental GeT_Move Algorithm

Until now, we have presented the Incremental GeT_Move which split the original cluster matrix into different equivalent blocks. The experiment results show that the algorithm is efficient. However, the disadvantage of this approach is that we do not know what is the optimal block size. To identify the optimal block sizes, different techniques can be applied, such as data sampling in which a sample of data is used to investigate the optimal block sizes. Even if this approach is appealing, extracting such a sample is very difficult.

To tackle this problem, we propose an innovative solution to dynamically assign blocks to Incremental GeT_Move. Before presenting the approach, we would like to propose the definition of a fully nested cluster matrix (resp. block) (Figure 9c) as follows.

Definition 9. *Fully nested cluster matrix (resp. block). An $n \times m$ 0-1 block $b$ is fully nested if for any two column $r_i$ and $r_{i+1}$, $r_i, r_{i+1} \in b$, we have $r_i \cap r_{i+1} = r_{i+1}$.*

We can consider that the $LCM$ is very efficient when applied on dense (resp. (fully) nested) datasets and blocks. Let $E$ be the universe of items, consisting of items $1, \ldots, n$. A subset $X$ of $E$ is called an itemset. In the LCM algorithm process on a common cluster matrix, for any $X$, we make the recursive call for $X[i]$ for each $i \in \{i(X)+1, \ldots, |E|\}$ because we do not know which $X[i]$ will be a closed itemset when $X$ is extended by adding $i$ to $X$. Meanwhile, for a fully nested cluster matrix, we know that only the recursive call for item $i = i(X)+1$ is valuable and the other recursive calls for each item $i \in \{i(X)+2, \ldots, |E|\}$ are useless. Note that $i(X)$ returns the last item of $X$.

Property 6. *Recursive Call. Given a fully nested cluster matrix nCM (resp. block), a universe of items E of nCM, an itemset X which is a subset of E. All the FCIs can be generated by making a recursive call of item $i = i(X)+1$.*

Proof. After construction, we have $\forall i \in E, O(i) \cap O(i+1) = O(i+1)$; thus, $O(i+1) \subseteq O(i)$. Additionally, $\forall i' \in \{i(X)+2, \ldots, |E|\}$ we need to make a recursive call for $X[i']$ and let assume that we obtain a frequent itemset $X \cup i' \cup X'$ with $X' \subseteq \{i(X)+3, \ldots, |E|\}$. We can consider that $O(i') \subseteq O\big(i(X)+1\big)$ and therefore $O(X \cup i' \cup X') = O\Big(X \cup \big(i(X)+1\big) \cup i' \cup X'\Big)$. Consequently, $X \cup i' \cup X'$ is not a FCI because $(X \cup i' \cup X') \subset \Big(X \cup \big(i(X)+1\big) \cup i' \cup X'\Big)$ and $O(X \cup i' \cup X') = O\Big(X \cup \big(i(X)+1\big) \cup i' \cup X'\Big)$. Furthermore, $\Big(X \cup \big(i(X)+1\big) \cup i' \cup X'\Big)$ can be generated by making a recursive call for $i(X)+1$. We can conclude that it is useless to make a recursive call for $\forall i' \in \{i(X)+2, \ldots, |E|\}$ and additionally, all FCIs can be generated only by making a recursive call for $i(X)+1$. □

By applying *Property 6*, we can consider that LCM is more efficient on a fully nested matrix because it reduces unnecessary recursive calls. Therefore, our goal is to retrieve fully nested blocks to improve the performance of Incremental GeT_Move. In order to reach this goal, we first apply the *nested and segment nested Greedy* algorithm [3] [31] to re-arrange the cluster matrix (Figure 9a) so that it now becomes a *nested cluster matrix* (Figure 9b). Then, we propose a sub-function *Nested Block Partition* (Figure 7-(4)) to dynamically split the nested cluster matrix into *fully nested blocks* (Figure 9c).

[3]http://www.aics-research.com/nestedness/

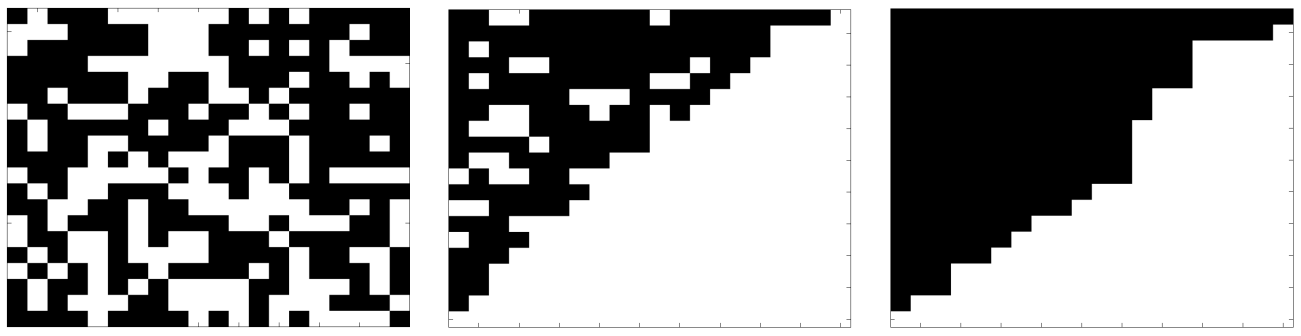

(a) (b) (c)

**Figure 9: Examples of non-nested , almost nested, fully nested datasets. Black = 1, white = 0. (a) Original, (b) Almost nested, (c) Fully nested.**

By following the definition 9 and scanning the nested cluster matrix from the beginning to the end, we are able to obtain all fully nested blocks. We start from the first column of nested cluster matrix, then we check the next column and if the nested condition is held then the block is expanded; otherwise, the block is set and we create a new block. Note that all small blocks containing only 1 column are merged into a sparse block $SpareB$. At the end, we obtain a set of fully nested blocks $NestedB$ and a sparse block $SpareB$. Finally, the Incremental GeT_Move is applied on $B = NestedB \cup SpareB$.

The pseudo code of Fully Nested Blocks Partition subfunction is described in *Algorithm 3.*

**Algorithm 3: Fully Nested Blocks Partition**

```
Input  : a nested cluster matrix CM_N
Output: a set of blocks B
begin
    B := ∅; NestedB := ∅; SpareB := ∅;
    foreach item i ∈ CM_N do
        if i ∩ (i + 1) = (i + 1) then
            NestedB := NestedB ∪ i;
        else
            NestedB := NestedB ∪ i;
            if |NestedB| ≤ 1 then
                SpareB.push_all(NestedB);
                NestedB := ∅
            else
                B := B ∪ NestedB;
                NestedB := ∅
    return B := B ∪ SpareB;
where the purpose SpareB.push_all(NestedB) function
is to put all items in NestedB to SpareB.
```

## 4.4 Spatio-Temporal Pattern Mining Algorithm Based on Explicit Combination of FCI Pairs

In real world applications (e.g. cars), object locations are continuously reported by using Global Positioning System (GPS). Therefore, new data is always available. Let us denote the new movement data as $(O_{DB}, T_{DB'})$. Naturally, it is cost-prohibitive and time consuming to execute Incremental GeT_Move (or GeT_Move) on the entire database (denoted $(O_{DB}, T_{DB} \cup T_{DB'})$) which is created by merging $(O_{DB}, T_{DB'})$ into the existing database $(O_{DB}, T_{DB})$. To tackle this issue, we provide an approach which efficiently combines the existing frequent closed itemsets $FCIs_{DB}$ with the new frequent closed itemsets $FCIs_{DB'}$, which are extracted from $DB'$, to obtain the final results $FCIs_{DB \cup DB'}$.

**Table 5: An example of FCI binary presentation.**

| | | $FCIs_{DB}$ | | | $FCIs_{DB'}$ | |
|---|---|---|---|---|---|---|
| $binary(FCI)$ | | $b(ci_1)$ | $b(ci_2)$ | | $b(ci'_1)$ | $b(ci'_2)$ |
| $O_{DB}$ | $o_1$ | 1 | 0 | | 1 | 0 |
| | $o_2$ | 1 | 0 | | 1 | 1 |
| | $o_3$ | 0 | 1 | | 0 | 1 |
| | $o_4$ | 0 | 1 | | 0 | 1 |

For instance, in Table 5, we have two sets of frequent closed itemsets $FCIs_{DB}$ and $FCIs_{DB'}$. Each FCI will be presented as a $|O_{DB}|$-bit binary numeral. For clarity sake, binary presentation of a FCI is used when applying binary operators (i.e. $\vee, \wedge$, etc). For instance, $b(ci_1) \vee b(ci'_1)$ is represented by $ci_1 \vee ci'_1$. On the other hand, they are considered as a list of items (resp. clusters) when set operators (i.e. $\cup, \cap, \subseteq, \in$, etc) are applied.

The principle function of our algorithm is to explicitly combine all pairs of $FCIs(ci, ci')$ to generate new FCIs. Let us assume that $ci \wedge ci' = \gamma$, $\gamma = ci \cup ci'$ is a FCI if $\sigma(\gamma)$ is larger than $\varepsilon$ and that there are no subsets of $O(ci)$, $O(ci')$ so that they are a superset of $O(\gamma)$. Here is an explicit combination of a pair of $FCIs(ci, ci')$:

PROPERTY 7. *Explicit Combination of a pair of FCIs. Given FCIs ci and ci' so that $ci \in FCIs_{DB}, ci' \in FCIs_{DB'}$, a . $ci \cup ci'$ is a FCI that belongs to $FCIs_{DB \cup DB'}$ if and only if:*

$$\begin{cases} \text{if } ci \wedge ci' = \gamma \text{ then} \\ (1): Size(\gamma) \geq \varepsilon. \\ (2): \nexists p : p \in FCIs_{DB}, O(\gamma) \subseteq O(p) \subseteq O(ci). \\ (3): \nexists p' : p' \in FCIs_{DB'}, O(\gamma) \subseteq O(p') \subseteq O(ci'). \end{cases} \quad (9)$$

*where $ci = \{c_{t_{a_1}}, c_{t_{a_2}}, \ldots, c_{t_{a_p}}\}$ and $ci' = \{c'_{t_{a_1}}, c'_{t_{a_2}}, \ldots, c'_{t_{a_p}}\}$, $Size(ci)$ returns the number of '1' in ci. Note that $Size(\gamma) = O(\gamma) = \sigma(\gamma)$.*

PROOF. After construction, we have $\nexists p : p \in FCIs_{DB}, O(\gamma) \subseteq O(p) \subseteq O(ci)$. We assume that $\exists i$ s.t. $i \in C_{DB}, O(\gamma) \subseteq i$ and $i \notin ci$ therefore $\exists p$ s.t. $p = \{\forall i | i \in C_{DB}, O(\gamma) \subseteq i, i \notin ci\} \cup ci, O(\gamma) \subseteq O(p)$. Consequently, $\forall i \in C_{DB}, O(\gamma) \subseteq i$ then $i \in p$ and therefore $p$ is a FCI and $p \in FCIs_{DB}$. This violates the assumption and therefore $\nexists i$ s.t. $i \in C_{DB}, O(\gamma) \subseteq i$ and $i \notin ci$ or $\forall i$ s.t. $i \in C_{DB}, O(\gamma) \subseteq i$ then $i \in ci$. Similarly, if $\nexists p' : p' \in FCIs_{DB'}, O(\gamma) \subseteq O(p') \subseteq O(ci')$ then $\forall i'$ s.t. $i' \in C_{DB'}, O(\gamma) \subseteq i'$ then $i' \in ci'$. Consequently, if $\forall i \in C_{DB \cup DB'}, O(\gamma) \subseteq i$ then $i \in ci \cup ci'$. Additionally, $Size(\gamma) = \sigma(\gamma) \geq \varepsilon$ and therefore $ci \cup ci'$ is a FCI and $ci \cup ci' \in FCIs_{DB \cup DB'}$. □

We can consider that if $ci \cup ci'$ is a FCI, they must respectively be the two longest FCIs which contain $O(\gamma)$ in $FCIs_{DB}$ and $FCIs_{DB'}$. $(O(\gamma), ci \cup ci')$ is a new FCI and it will be stored in a set of new frequent closed itemsets, named $FCIs_{new}$. To efficiently make all combinations, we first partition $FCIs_{DB}, FCIs_{DB'}$ and $FCIs_{new}$ into different partitions in terms of support so that the FCIs, that have

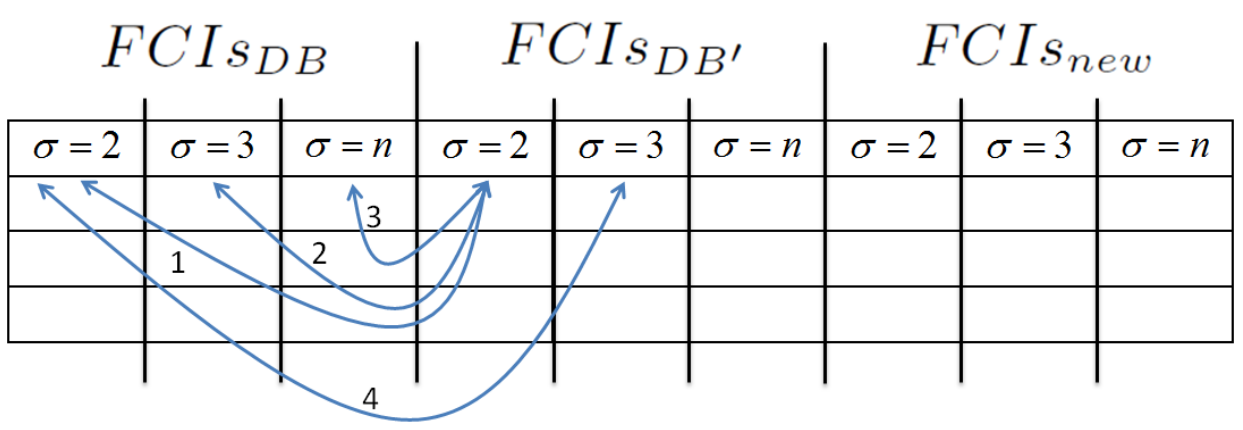

**Figure 10: An example of the explicit combination of pairs of FCIs-based approach.**

the same support value, will be in the same partition (Figure 10). Secondly, partitions are combined from the smallest support values (resp. longest FCIs) to the largest ones (resp. shortest FCIs). New FCIs will be added into the right partition in $FCIs_{new}$. By using this approach, it is guaranteed that the first time there is $ci \wedge ci' = \gamma, Size(\gamma) \geq \varepsilon$ then $ci \cup ci'$ is a new FCI because they are the two longest FCIs which contain $O(\gamma)$. Therefore, we just ignore the later combinations which return $\gamma$ as the result. Furthermore, to ensure that $\gamma$ already exists in $FCIs_{new}$ or not, we only need to check items in the $FCIs_{new}$ partition whose support value is equal to $Size(\gamma)$. We can consider that by partitioning $FCIs_{DB}, FCIs_{DB'}$ and $FCIs_{new}$, the process is much improved. Additionally, we also propose a pruning rule to speed up the approach by ending the combination running of a FCI $ci'$ as follows:

Lemma 4. *The combination running of $ci'$ is stopped if:*

$$\exists ci \in FCIs_{DB} \text{ s.t. } ci \wedge ci' = ci', ci \cup ci' \text{ is a FCI.} \quad (10)$$

Proof. Assume that $\exists \Upsilon : \Upsilon \in FCIs_{DB}, \sigma(\Upsilon) \geq \sigma(ci), \Upsilon \wedge ci' = ci'$. If $O(ci) \subseteq O(\Upsilon)$ then we have $ci \in FCIs_{DB}, O(ci') \subseteq O(ci) \subseteq O(\Upsilon)$ and this violates the condition 2 in *Property 7*, therefore $\Upsilon \cup ci'$ is not a FCI. If $O(ci) \nsubseteq O(\Upsilon)$ then $\exists i \in C_{DB}$ s.t. $O(ci') \subseteq i$ and $i \notin \Upsilon$. Furthermore, $\exists p : p = \{\forall i | i \in C_{DB}, O(ci') \subseteq i, i \notin \Upsilon\} \cup \Upsilon$. So, $\forall i, i \in C_{DB}, O(ci') \subseteq i$ then $i \in p$ and therefore $p$ is a FCI and $p \in FCIs_{DB}$. Additionally, $O(ci') \subseteq O(p) \subseteq O(\Upsilon)$. This violates the condition 2 in *Property 7*, therefore $\Upsilon \cup ci'$ is not a FCI. Consequently, we can conclude that $\nexists \Upsilon$ s.t. $\Upsilon \in FCIs_{DB}, \sigma(\Upsilon) \geq \sigma(ci), \Upsilon \wedge ci' = ci'$ and $\Upsilon \cup ci'$ is a FCI. Therefore, we do not need to continue the combination running of $ci'$. □

Similar to lemma 4, in the explicit combination process, $ci$ will be disactivated for further combinations when there is a $ci'$ so that $ci \wedge ci' = ci$ and $ci \cup ci'$ is a FCI. After generating all new FCIs in $FCIs_{new}$, the final results $FCIs_{DB \cup DB'}$ is created by collecting FCIs in $FCIs_{DB}, FCIs_{DB'}, FCIs_{new}$. In this step, some of them will be discarded such that:

Property 8. *Discarded FCIs in $FCIs_{DB \cup DB'}$ creating step. All the FCIs which satisfy the following conditions will not be selected as a FCIs in the final results.*

$$\begin{cases} (1): \forall ci \in FCIs_{DB},\ \text{if } \exists ci' \in FCIs_{DB'} \text{ s.t.} \\ ci \wedge ci' = ci \text{ then } ci \text{ will not be selected.} \\ (2): \forall ci' \in FCIs_{DB'},\ \text{if } \exists ci \in FCIs_{DB} \text{ s.t.} \\ ci \wedge ci' = ci' \text{ then } ci' \text{ will not be selected.} \end{cases} \quad (11)$$

Note that during the explicit combination step, the FCIs which will not be selected for the final results are removed by applying the conditions in Property 8. It means that we only add all suitable FCIs into $FCIs_{DB \cup DB'}$ and therefore it is optimized and much less costly. In the worst case scenario, the complexity of explicit combination of pairs of FCIs step is $O(|FCIs_{DB}| \times |FCIs_{DB'}| \times \frac{|FCIs_{new}|}{\#partitions(FCIs_{new})})$. Naturally, $T_{DB'}$ is much smaller than $T_{DB}$ and therefore $FCIs_{DB'}, FCIs_{new}$ are very small compare to $FCIs_{DB}$. Consequently, the process can be potentially greatly improved when compared to executing the Incremental GeT_Move on the entire database $(O_{DB}, T_{DB \cup DB'})$.

The pseudo code of *Explicit Combination of Pairs of FCIs-based Spatio-Temporal Pattern Mining Algorithm* is described in *Algorithm 4*.

**Algorithm 4: Explicit Combination of Pairs of FCIs-based Spatio-Temporal Pattern Mining Algorithm**

**Input** : a set of FCIs $FCIs_{DB}$, Occurrence sets $K$, int $\varepsilon$, int $min_t$, double $\theta$, set of Occurrence sets (blocks) $B'$, int $min_c$, double $min_{wei}$

1 **begin**
2 $FCIs_{DB'} := \emptyset; FCIs_{new} := \emptyset; FCIs_{DB \cup DB'} := \emptyset;$
3 $FCIs_{DB'} :=$ Incremental GeT_Move* $(K, \varepsilon, min_t, CI, \theta, B', min_c, min_{wei});$
4 **foreach** *partition* $P' \in FCIs_{DB'}$ **do**
5 **foreach** *FCI* $ci' \in P'$ **do**
6 **foreach** *partition* $P \in FCIs_{DB}$ **do**
7 **foreach** *FCI* $ci \in P$ **do**
8 $\gamma := ci \wedge ci';$
9 **if** $Size(\gamma) \geq \varepsilon$ *and* $FCIs_{new}.notContain(\gamma, Size(\gamma))$ **then**
10 $\gamma := ci \cup ci';$
11 $FCIs_{new}.add(\gamma, Size(\gamma));$
12 **if** $\gamma = ci$ **then**
13 $FCIs_{DB}.remove(ci);$
14 **if** $\gamma = ci'$ **then**
15 $FCIs_{DB}.remove(ci');$
16 **go to** line 5;
17 $FCIs_{DB \cup DB'} := FCIs_{DB} \cup FCIs_{DB'} \cup FCIs_{new};$
**foreach** *FCI* $X \in FCIs_{DB \cup DB'}$ **do**
18 **PatternMining**$(X, min_t)$; /*$X$ is a pattern?*/
19 Where: Incremental GeT_Move* is an Incremental GeT_Move without *PatternMining* sub-function, $FCIs_{new}.notContain(\gamma, Size(\gamma))$ returns true if there does not exists $\gamma$ in partition which has the support value is $Size(\gamma)$.

## 5. EXPERIMENTAL RESULTS

A comprehensive performance study has been conducted on real datasets and synthetic datasets. All the algorithms are implemented in C++, and all the experiments are carried out on a 2.8GHz Intel Core i7 system with 4GB Memory. The system runs Ubuntu 11.10 and g++ version 4.6.1.

The implementation of our proposed algorithms mining is also integrated in our demonstration system and it is public online[4]. As in [6], the two following datasets[5] have been used during experiments: *Swainsoni dataset* includes 43 objects evolving over time and 764 different timestamps. The dataset was generated from July 1995 to June 1998. *Buffalo dataset* concerns 165 buffalos and the tracking time from year 2000 to year 2006. The original data has 26610 reported locations and 3001 timestamps.

Similar to [6, 3, 16], we first use linear interpolation to fill in the missing data. For study purposes, we needed the objects to stay together for at least $min_t$ timestamps. As [6, 3, 16], DBScan [5] ($MinPts = 2, Eps = 0.001$) is applied to generate clusters at each timestamp.

## 5.1 Effectiveness

We proved that mining spatio-temporal patterns can be similarly mapped into itemsets mining issue. Therefore, in theoretical way, our approaches can provides the correct results. Experimentally, we do a further comparison, we first obtain the spatio-temporal patterns by employing traditional algorithms such as, $CMC, CuTS^{*}$[6] (convoy mining), $ObjectGrowth$ (closed swarm mining) as well as our approaches. To apply our algorithms, we split cluster matrix into blocks such as each block $b$ contains 25 consecutive timestamps. Additionally, to retrieve all the spatio-temporal patterns, in the reported experiments, the default value of $\varepsilon$ is set to 2 (two objects can form a pattern), $min_t$ is 1. Note that the default values are the hardest conditions for examining the algorithms. Then in the following we mainly focus on different values of $min_t$ in order to obtain different sets of convoys, closed swarms and group patterns. Note that for group patterns, $min_c$ is 1 and $min_{wei}$ is 0.

The results show that our proposed approaches obtain the same results compared to the traditional algorithms. An example of patterns is illustrated in Figure 11. For instance, see Figure 11a, a closed swarm is discovered within a frequent closed itemset. Furthermore, from the itemset, a convoy and a group pattern are also extracted (i.e. Figure 11b, 11c).

## 5.2 Efficiency

### 5.2.1 Incremental GeT_Move and GeT_Move Efficiency

To show the efficiency of our algorithms, we also generate larger synthetic datasets using Brinkhoff's network[7]-based generator of moving objects as in [6]. We generate 500 objects ($|O_{DB}| = 500$) for $10^4$ timestamps ($|T_{DB}| = 10^4$) using the generator's default map with low moving speed (250). There are $5 \times 10^6$ points in total. DBScan ($MinPts = 3, Eps = 300$) is applied to obtain clusters for each timestamp.

In the efficiency comparison, we employ $CMC, CuTS^{*}$ and $ObjectGrowth$. Note that, in [6], $ObjectGrowth$ outperforms $VG{-}Growth$ [21] (a group patterns mining algorithm) in terms of performance and therefore we will only consider $ObjectGrowth$ and not both. Note that GeT_Move and Incremental GeT_Move extracted closed swarms, convoys and group patterns meanwhile $CMC, CuTS^{*}$ only extracted convoys and $ObjectGrowth$ extracted closed swarms.

**Efficiency w.r.t.** $\varepsilon, min_t$. Figure 12a, 13a, 14a show running time w.r.t. $\varepsilon$. It is clear that our approaches outperform other algorithms. ObjectGrowth is the lowest one and the main reason is that with low $min_t$ (default $min_t = 1$), the *Apriori Pruning* rule (the most efficient pruning rule) is no longer effective. Therefore, the search space is greatly enlarged ($2^{|O_{DB}|}$ in the worst case). Additionally, there is no pruning rule for $\varepsilon$ and therefore the change of $\varepsilon$ does not directly affect the running time of ObjectGrowth. A little bit further, GeT_Move is lower than Incremental GeT_Move.

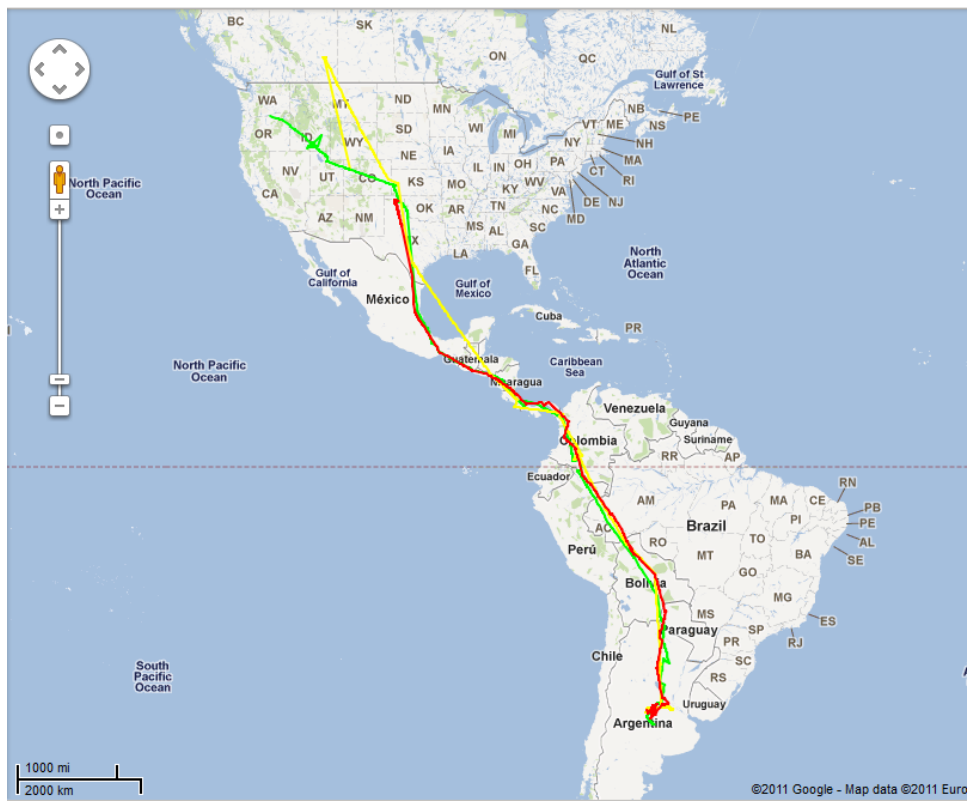

(a) One of discovered closed swarms.

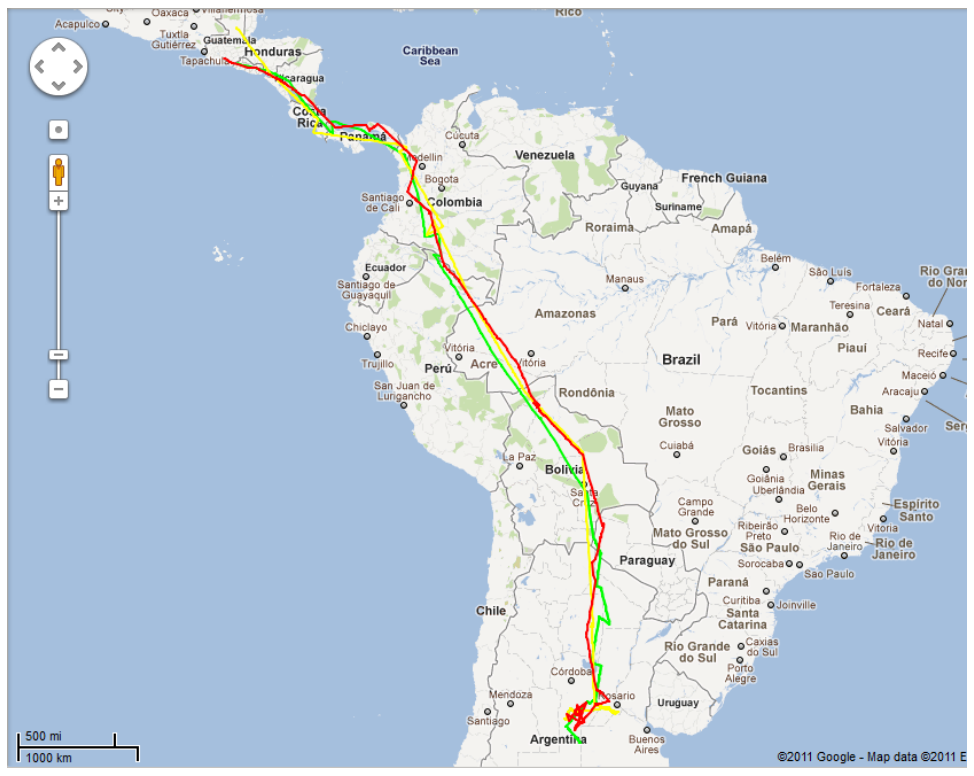

(b) One of discovered convoys.

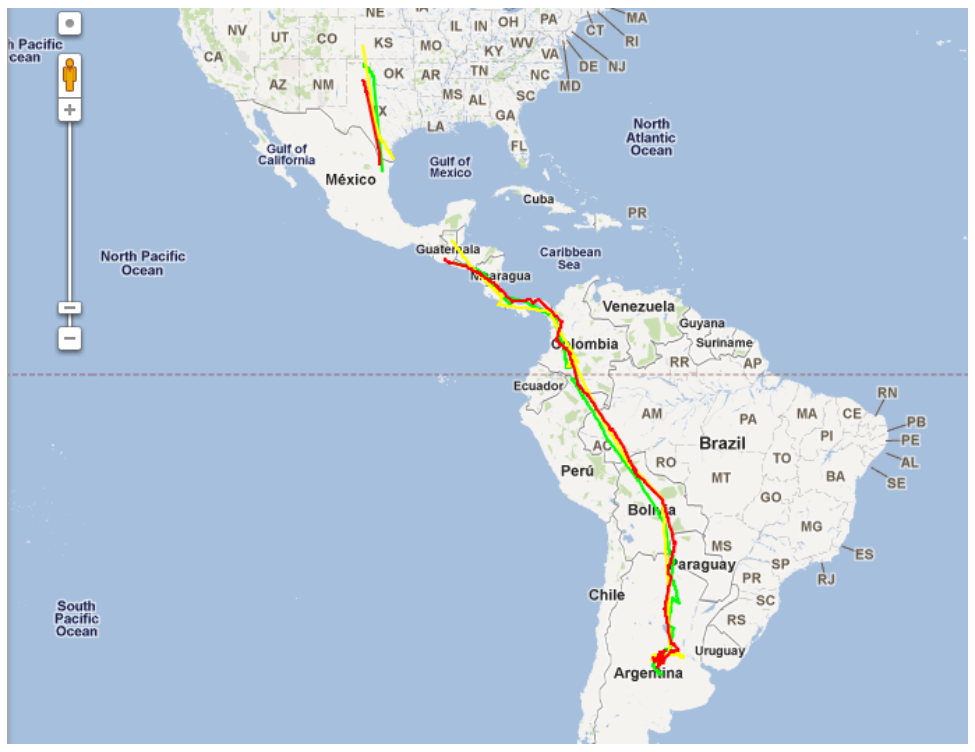

(c) One of discovered group patterns.

**Figure 11: An example of patterns discovered from Swainsoni dataset.**

---

[4] www.lirmm.fr/~phan/index.jsp

[5] http://www.movebank.org

[6] The source code of $CMC, CuTS^{*}$ is available at http://lsirpeople.epfl.ch/jeung/source_codes.htm

[7] http://iapg.jade-hs.de/personen/brinkhoff/generator/

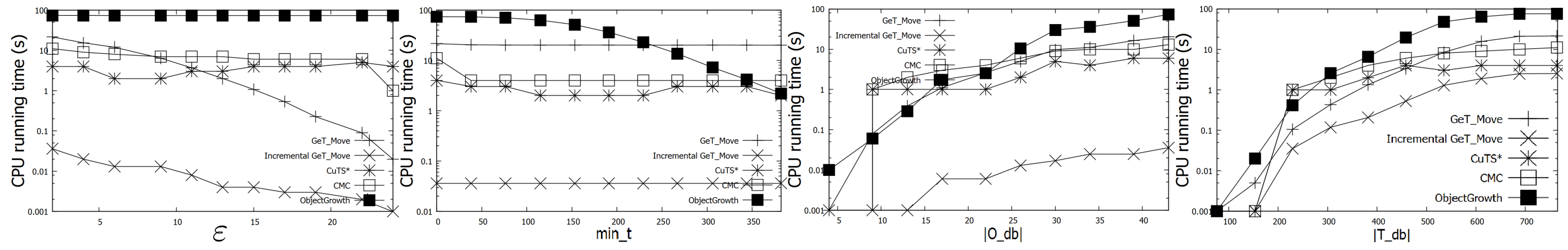

(a) Running time w.r.t. $\varepsilon$ (b) Running time w.r.t. $min_t$ (c) Running time w.r.t. $|O_{DB}|$ (d) Running time w.r.t. $|T_{DB}|$

**Figure 12: Running time on Swainsoni Dataset.**

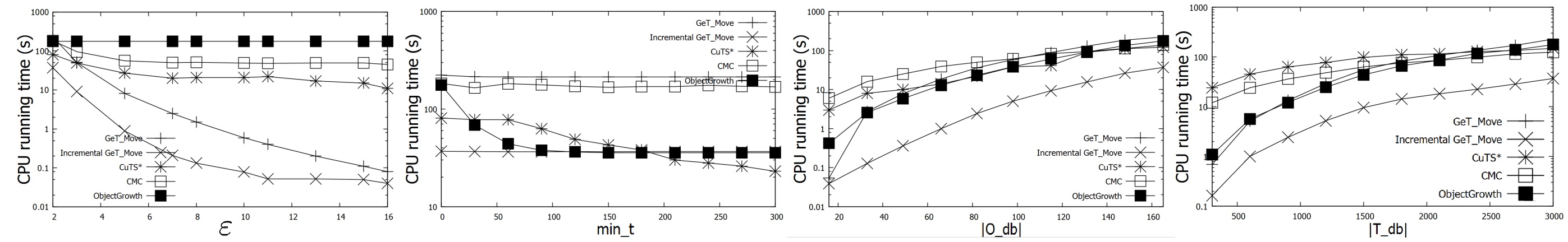

(a) Running time w.r.t. $\varepsilon$ (b) Running time w.r.t. $min_t$ (c) Running time w.r.t. $|O_{DB}|$ (d) Running time w.r.t. $|T_{DB}|$

**Figure 13: Running time on Buffalo Dataset.**

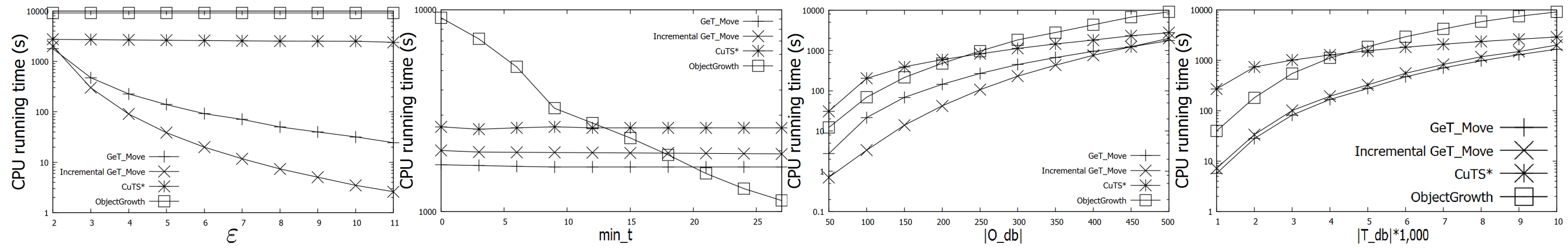

(a) Running time w.r.t. $\varepsilon$ (b) Running time w.r.t. $min_t$ (c) Running time w.r.t. $|O_{DB}|$ (d) Running time w.r.t. $|T_{DB}|$

**Figure 14: Running time on Synthetic Dataset.**

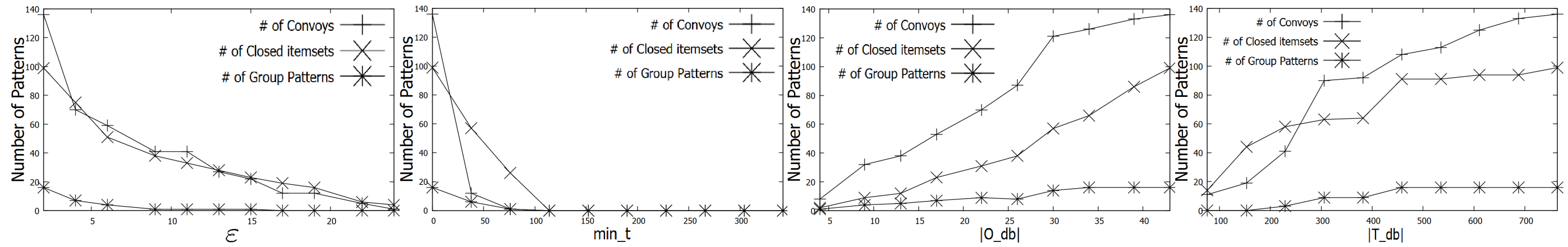

(a) # of patterns w.r.t. $\varepsilon$ (b) # of patterns w.r.t. $min_t$ (c) # of patterns w.r.t. $|O_{DB}|$ (d) # of patterns w.r.t. $|T_{DB}|$

**Figure 15: Number of patterns on Swainsoni Dataset. Note that # of frequent closed itemsets is equal to # of closed swarms.**

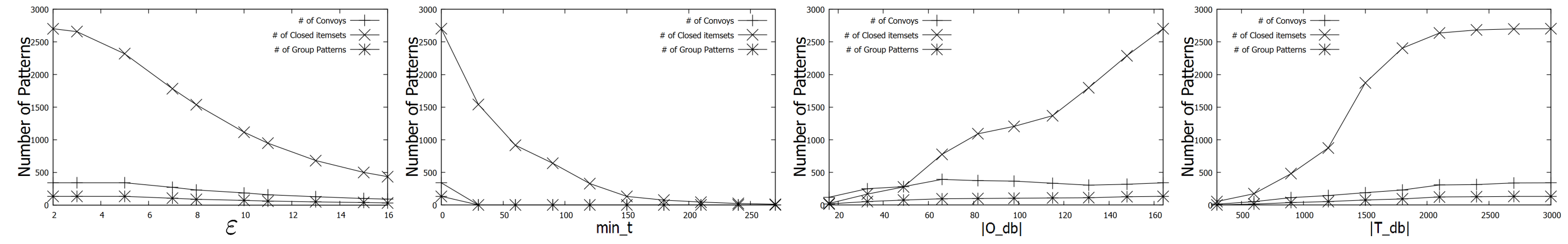

(a) # of patterns w.r.t. $\varepsilon$ (b) # of patterns w.r.t. $min_t$ (c) # of patterns w.r.t. $|O_{DB}|$ (d) # of patterns w.r.t. $|T_{DB}|$

**Figure 16: Number of patterns on Buffalo Dataset. Note that # of frequent closed itemsets is equal to # of closed swarms.**

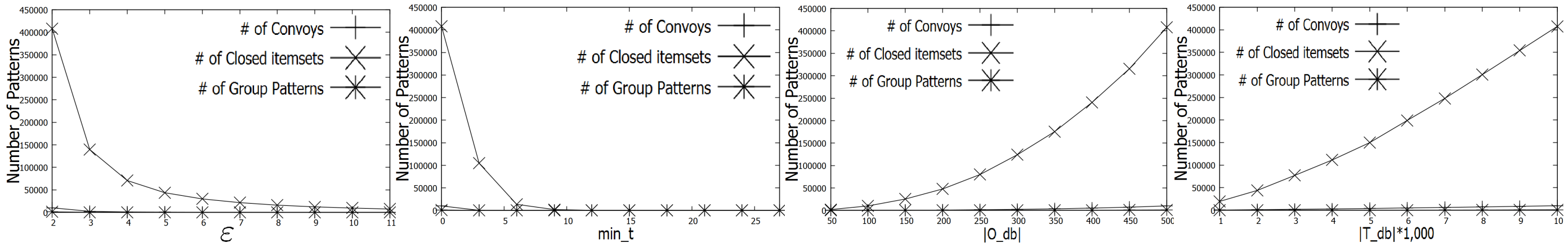

(a) # of patterns w.r.t. $\varepsilon$ (b) # of patterns w.r.t. $min_t$ (c) # of patterns w.r.t. $|O_{DB}|$ (d) # of patterns w.r.t. $|T_{DB}|$

**Figure 17: Number of patterns on Synthetic Dataset. Note that # of frequent closed itemsets is equal to # of closed swarms.**

The main reason is that GeT_Move has to proccess with large number of items and long itemsets. Meanwhile, thanks to blocks, the number of items is greatly reduced and itemsets are not long as the ones in GeT_Move.

Figure 12b, 13b, 14b show running time w.r.t. $min_t$. In almost all cases, our approaches outperform other algorithms. See Figure 13b, 14b, with low $min_t$, our algorithm is much faster than the others. However, when $min_t$ is higher ($min_t > 200$ in Figure 13b, $min_t > 20$ in Figure 14b) our algorithms take more time than CuTS* and ObjectGrowth. This is because with high value of $min_t$, the number of patterns is significantly reduced (Figure 15b, 16b, 17b) (i.e. no extracted convoy when $min_t > 100$ (resp. $min_t > 200$, $min_t > 10$), Figure 15b (resp. 16b, 17b)) and therefore CuTS* and ObjectGrowth is faster. Meanwhile, GeT_Move and Incremental GeT_Move have to work with frequent closed itemsets.

**Efficiency w.r.t.** $|O_{DB}|, |T_{DB}|$. Figure 12c-d, Figure 13c-d, Figure 14c-d show the running time when varying $|O_{DB}|$ and $|T_{DB}|$ respectively. In all figures, Incremental GeT_Move outperforms other algorithms. However, with synthetic data (Figure 14d) and lowest values of $\varepsilon = 2$ and $min_t = 1$, GeT_Move is a little bit faster than Incremental GeT_Move. This is the clue to the fact that Incremental GeT_Move does not have any information to obtain the better partitions (blocks).

**Scalability w.r.t.** $\varepsilon$. We can consider that the running time of algorithms does not change significantly when varied $min_t, |O_{DB}|, |T_{DB}|$ in synthetic data (Figures 14). However, they are quite different when varying $\varepsilon$ (default $min_t = 1$). Therefore, we generate another large synthetic data to test the scalability of algorithms on $\varepsilon$. The dataset includes 50,000 objects moving during 10,000 timestamps and it contains 500 million locations in total. The executions of CMC and CuTS* stop due to a lack of memory capacity after processing 300 milion locations. Additionally, ObjectGrowth can not provide the results after 1day running. The main reason is that with low $min_t$ ($= 1$), the search space is significant larger ($\approx 2^{50,000}$). Meanwhile, thanks to the LCM approach, our algorithms can provide the results within hours (Figure 18).

**Efficiency w.r.t.** *Block-size.* To investigate the optimal value of block-size, we examine Incremental GeT_Move by using the default values of $\epsilon, min_t$ with different block-size values on real datasets and synthetic dataset ($|O_{DB}| = 500, |T_{DB}| = 1,000$). The optimal block-size range can be from 20 to 30 timestamps within which Incremental GeT_Move obtains the best performance for all the datasets (Figure 19). The main reason is that objects tend to move together in suitable short interval (from 20 to 30 timestamps). Therefore, by setting block-size in this range, the data is efficiently compressed into FCIs. Meanwhile, with larger block-size values, the objects' movements are quite different; therefore, the data compressing is not so efficient. Regarding to small block-size values (5-15), we have to face up to a large number of blocks so that the process is slowed down. In the previous experiments, block-size is set to 25.

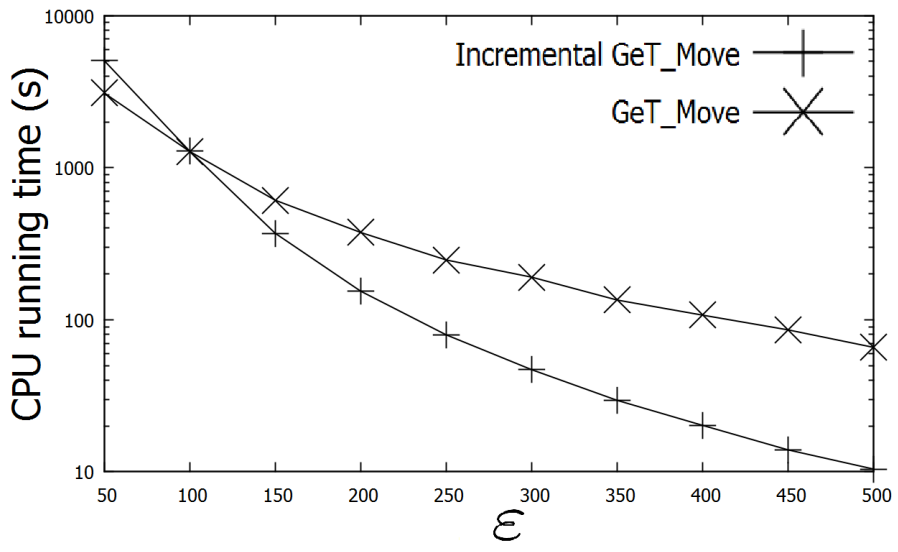

**Figure 18: Running time w.r.t $\varepsilon$ on large Synthetic Dataset.**

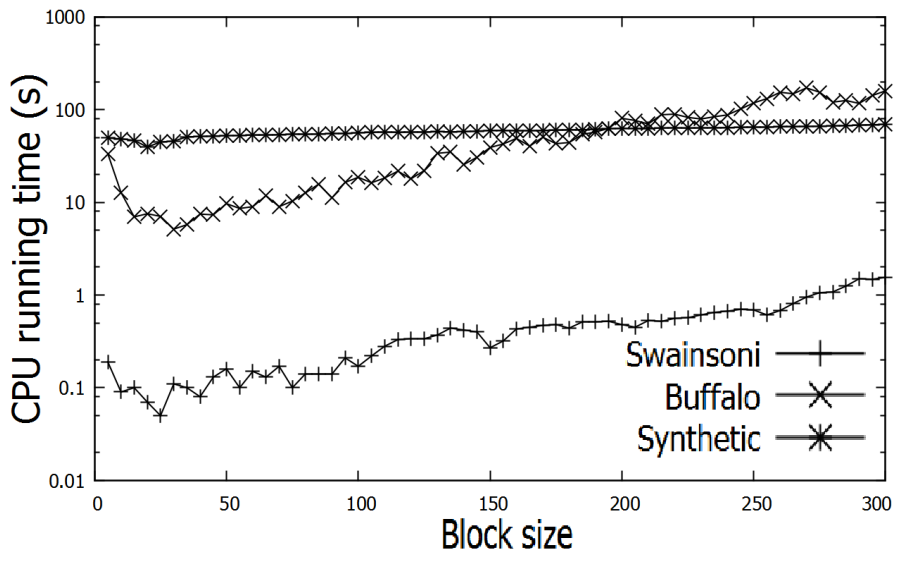

**Figure 19: Running time w.r.t block size.**

### 5.2.2 *Parameter Free Incremental GeT_Move Efficiency*

The experimental results show that, so far, Incremental GeT_Move and GeT_Move outperform other algorithms. Additionally, our algorithms can work with low $\varepsilon$ and $min_t$ values. In this section, we perform another experiment to examine the efficiency of the Parameter Free Incremental GeT_Move algorithm. In this experiment, we compare performances of four algorithms: 1) Parameter free Incremental GeT_Move, named Nested Incremental GeT_Move, 2) Nested GeT_Move which is the application of GeT_Move on nested cluster matrix $CM_N$, 3) Incremental GeT_Move which is executed with the optimal block size values on original cluster matrix $CM$, 4) GeT_Move which is applied on original cluster matrix $CM$.

**Efficient w.r.t.** *Real datasets.* Figure 21, 22 show that Nested Incremental GeT_Move (resp. Parameter Free Incremental GeT_Move) greatly outperforms the other algorithms. It is due to the better performance of LCM algorithm on nested cluster matrix (resp. fully nested blocks) compared to the original cluster matrix. Essentially, with the nested cluster matrix, the number of combinations of closed itemsets $X$ and items $i$ to ensure the closeness is greatly reduced. Therefore, the performance of the LCM algorithm is greatly improved. The fact is Nested GeT_Move is always better than GeT_Move (Figure 21, 22, 23). Additionally, the Swainsoni and Buffalo datasets contain many fully nested blocks (Table 6 and Figure 20). Consequently, the Nested Incremental GeT_Move is more efficient than the other algorithms.

**Efficient w.r.t.** *Synthetic dataset.* We can consider that Nested Incremental GeT_Move is quite similar to Nested GeT_Move (Figure 23). The main reasons are that: 1) Synthetic data is very sparse, 2) there are few fully nested blocks, 3) the nested blocks contain a very small number of items (i.e. 0.1% matrix fill by '1' and only 8 fully nested blocks which average length is 2, see Table 6 and Figure 20e-f). Therefore, the processing time of nested blocks is quite short. Meanwhile, there is a large nested sparse block which is the main partition that need to be processed by both Nested Incremental GeT_Move and Nested GeT_Move.

Additionally, thanks to the nested sparse block, the performance of LCM is improved a lot. Therefore, Nested Incremental GeT_Move and Nested GeT_Move are better than the others in most of cases. Exceptionally, with small number of objects $|O_{DB}|$ (i.e. $|O_{DB}| = 50$, Figure 23c) or high $\varepsilon$ (i.e. $\varepsilon \geq 9$, Figure 23a), Incremental GeT_Move is slightly better than Nested Incremental GeT_Move and Nested GeT_Move. The main reason is that Incremental GeT_Move splits the cluster matrix $CM$ into different small blocks within which there are a small number of items and FCIs which means that the computation cost is reduced. On the other hand, Nested Incremental GeT_Move and Nested GeT_Move need to work with a large nested sparse block.

#### 5.2.3 Spatio-Temporal Pattern Mining Algorithm Based on Explicit Combination of FCI Pairs

In this section, an experiment is designed to examine the spatio-temporal pattern mining algorithm based on explicit combination of FCI pairs and to identify when we should update the database. We first use half of Swainsoni, Buffalo and Synthetic datasets as a $DB$. Then the other half is used to generate $DB'$ which is increased step by step up to the maximum size (Figure 24). In this experiment, Incremental GeT_Move is employed to extract FCIs from $DB$ and $DB'$.

For real datasets (Swainsoni and Buffalo), the explicit combination algorithm is more efficient than the Incremental GeT_Move in all cases (Figure 24a, b). This is because we already have $FCIs_{DB}$ and therefore we only need to extract $FCIs_{DB'}$ and then combine $FCIs_{DB}$ and $FCIs_{DB'}$. Additionally, the Swainsoni and Buffalo are sufficiently dense (i.e. 17.8% and 7.2% with large number of fully nested blocks, see Table 6) so that the numbers of FCIs in $FCIs_{DB}$ and $FCIs_{DB'}$ are not huge. Consequently, the number of combinations is reduced and thus the algorithm is more efficient. In Figure 24a, b, we can consider that the running time of the explicit combination algorithm significantly changes when $|T_{DB'}| > 15\%|T_{DB}|$. This means that it is better to update the database when $|T_{DB'}| < 15\%|T_{DB}|$.

**Table 6: Fully nested blocks on datasets.**

| Dataset | Matrix fill | #Nested blocks | avg.length |
|---|---|---|---|
| Swainsoni | 17.8% | 102 | 4.52 |
| Buffalo | 7.2% | 602 | 2.894 |
| Synthetic | 0.1% | 8 | 2.00 |

For the synthetic dataset, the explicit combination algorithm is only efficient on small $DB'$ (i.e. $|T_{DB'}| < 20\%|T_{DB}|$, Figure 24c) because the dataset is very sparse. In fact, the number of FCIs in $FCIs_{DB'}$ is enlarged when the size of $DB'$ increases. Thus, the explicit combination algorithm is not efficient because of the huge number of combinations.

Overall, we can consider that the explicit combination algorithm obtains good efficiency when $T_{DB'}$ is smaller than 15% of $T_{DB}$.

To summarize, Incremental GeT_Move and GeT_Move outperform the other algorithms. Additionally, our algorithms can work with low values of $\varepsilon$ and $min_t$. To reach optimal efficiency, we propose a parameter free Incremental GeT_Move (resp. Nested Incremental GeT_Move) which dynamically assigns fully nested blocks for the algorithm from the nested cluster matrix. The experimental results show that the efficiency is greatly improved with the Nested Incremental GeT_Move and Nested GeT_Move. Furthermore, by storing FCIs in a closed itemset database (see Figure 7), it is possible to reuse them whenever new object movements arrive. The experimental results show that it is better to update the database when $T_{DB'}$ is smaller than 15% of $T_{DB}$ by applying the explicit combination algorithm.

## 6. CONCLUSION AND DISCUSSION

In this paper, we propose a (parameter free) unifying incremental approaches to automatically extract different kinds of spatio-temporal patterns by applying frequent closed itemset mining techniques. Their effectiveness and efficiency have been evaluated by using real and synthetic datasets. Experiments show that our approaches outperform traditional ones.

Another issue we plan to address is how to take into account the arrival of new objects which were not available for the first extraction. Now, as we have seen, we can store the results (resp. FCIs) to improve the process when new object movements arrive. In this approach we take the hypothesis is that the number of objects remains the same. However in some applications these objects could be different.

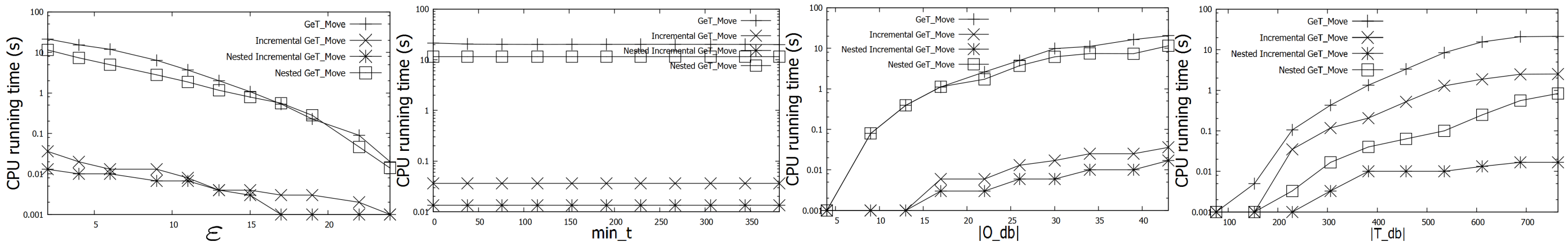

(a) Running time w.r.t. $\varepsilon$ (b) Running time w.r.t. $min_t$ (c) Running time w.r.t. $|O_{DB}|$ (d) Running time w.r.t. $|T_{DB}|$

**Figure 21: Running time on Swainsoni Dataset.**

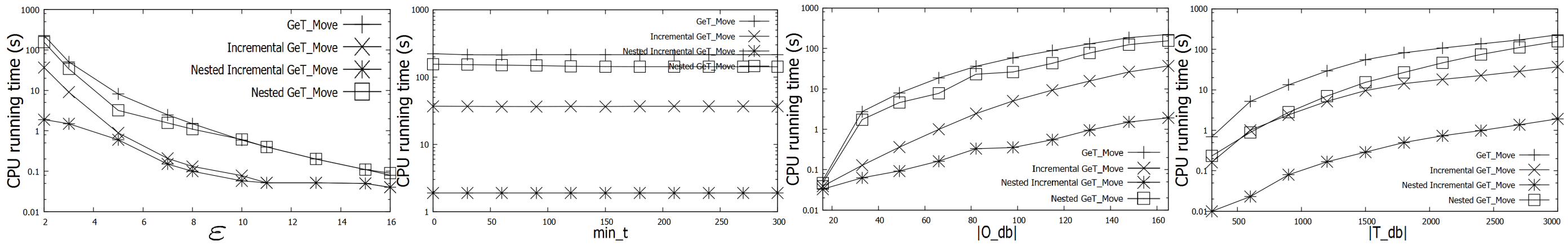

(a) Running time w.r.t. $\varepsilon$ (b) Running time w.r.t. $min_t$ (c) Running time w.r.t. $|O_{DB}|$ (d) Running time w.r.t. $|T_{DB}|$

**Figure 22: Running time on Buffalo Dataset.**

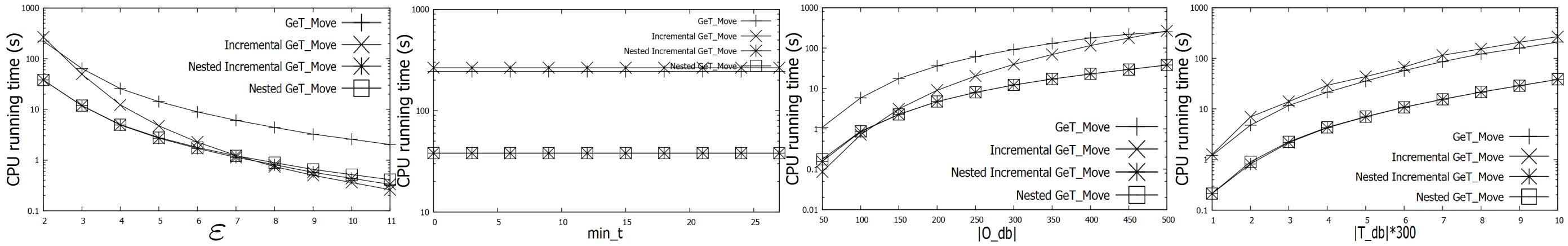

(a) Running time w.r.t. $\varepsilon$ (b) Running time w.r.t. $min_t$ (c) Running time w.r.t. $|O_{DB}|$ (d) Running time w.r.t. $|T_{DB}|$

**Figure 23: Running time on Synthetic Dataset.**

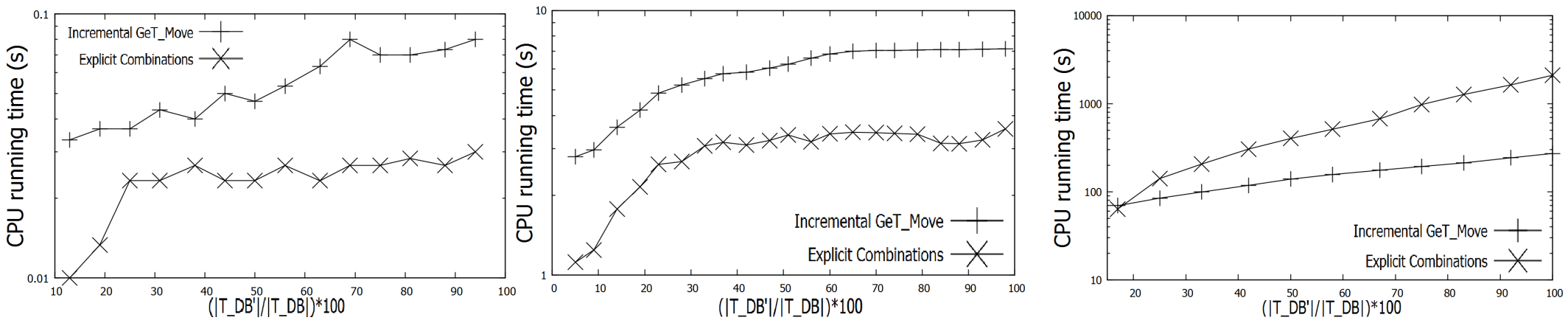

(a) Running time on Swainsoni (b) Running time on Buffalo (c) Running time on Synthetic data

**Figure 24: Explicit combinations algorithm efficiency.**

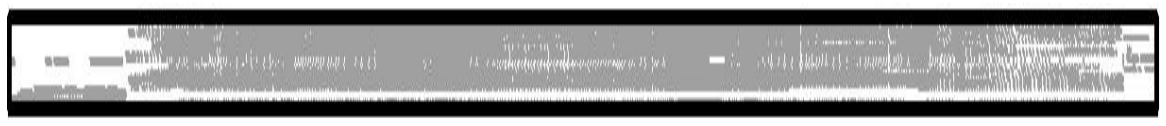

(a) Original Swainsoni cluster matrix.

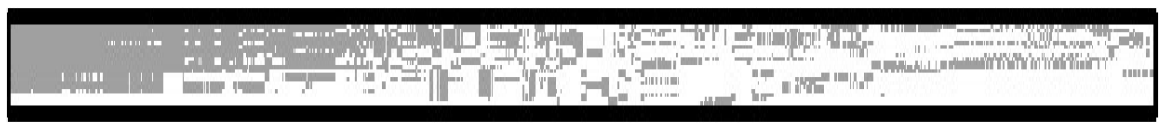

(b) Nested Swainsoni cluster matrix.

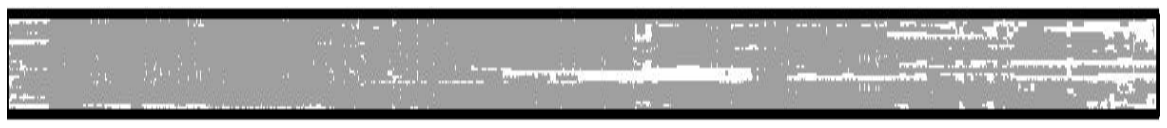

(c) Original Buffalo cluster matrix.

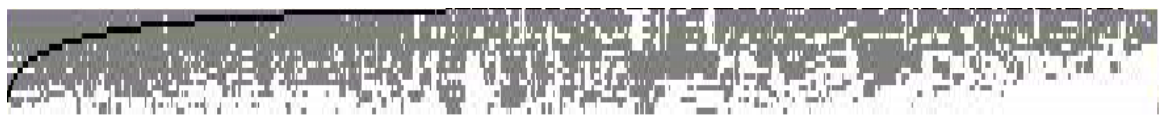

(d) Nested Buffalo cluster matrix.

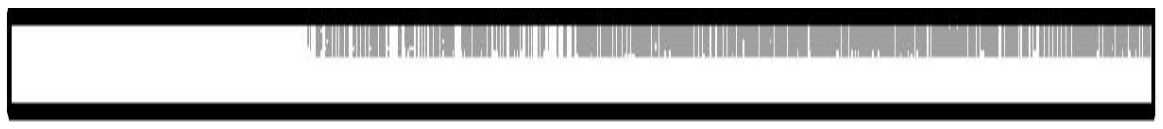

(e) Original Synthetic cluster matrix.

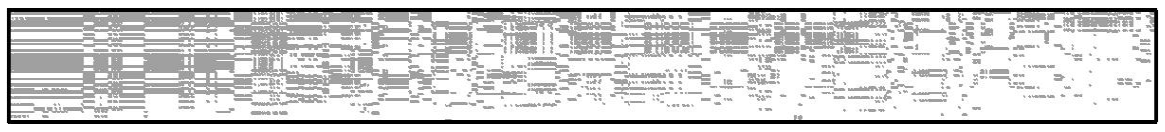

(f) Nested Synthetic cluster matrix.

**Figure 20: Original cluster matrices and nested cluster matrices.**